\documentclass[twocolumn,secnumarabic,german,aps,amsmath,amssymb]{revtex4}
\usepackage{graphicx,bm,float}
\usepackage{color}
\usepackage{amsmath}
\usepackage{amssymb}
\usepackage{mathrsfs} 

\begin{document}%
\title{Brittle yielding in supercooled liquids below the critical temperature of 
mode coupling theory}
\author {Konstantin Lamp, Niklas K\"uchler, and J\"urgen Horbach}
\affiliation{Institut f\"ur Theoretische Physik II: Weiche Materie, 
Heinrich-Heine-Universit\"at D\"usseldorf, Universit\"atsstra\ss e 1, 
40225 D\"usseldorf, Germany}
\begin{abstract}
Molecular Dynamics (MD) computer simulations of a polydisperse
soft-sphere model under shear are presented. Starting point for
these simulations are deeply supercooled samples far below the
critical temperature, $T_c$, of mode coupling theory. These samples
are fully equilibrated with the aid of the swap Monte Carlo technique.
For states below $T_c$, we identify a life time $\tau_{\rm lt}$
that measures the time scale on which the system can be considered
as an amorphous solid. The temperature dependence of $\tau_{\rm
lt}$ can be well described by an Arrhenius law. The existence of
transient amorphous solid states below $T_c$ is associated with the
possibility of brittle yielding, as manifested by a sharp stress
drop in the stress-strain relation and shear banding.  We show that
brittle yielding requires on the one hand low shear rates and on
the other hand, the time scale corresponding to the inverse shear
rate has to be smaller or of the order of $\tau_{\rm lt}$.  Both
conditions can be only met for large life time $\tau_{\rm lt}$,
i.e.~for states far below $T_c$.
\end{abstract}
%

\maketitle

\section{Introduction}
\label{sec1}
Glassforming liquids exhibit a dramatic slowing down of their
dynamics with decreasing temperature $T$. Important insight on the
origin of this slowing down has been given by the mode coupling
theory (MCT) of the glass transition \cite{goetze2009}. This theory
predicts a divergence of the structural relaxation time of the
liquid when decreasing $T$ towards a critical temperature
$T_c$.  At $T_c$, a transition from an ergodic liquid state to a
non-ergodic amorphous solid state occurs. The order parameter of
this transition is associated with the localization of each particle
in the cage that is formed by neighboring particles.  Thus, in the
framework of MCT, the glass transition can be seen as a localization
transition where, approaching the transition from temperatures
$T<T_c$, i.e.~from below, the critical temperature $T_c$ marks the
stability limit of the amorphous solid. At $T_c$, the length scale
$\xi$, that measures the localization of the particles in their
cages, reaches a critical value such that the amorphous solid state
cannot be stable anymore (note the analogy with the Lindemann
criterion for crystalline solids \cite{solyom2007}).

In real glassforming systems, a transition, as predicted by MCT,
is not observed. However, using the predictions of MCT, a critical
temperature $T_c$ can be identified around which the dynamics of
the supercooled liquid gradually changes from a liquid-like to a
solid-like dynamics \cite{cavagna2009}. As a consequence, far below
$T_c$, the supercooled liquid can be found in the state of an
amorphous solid, albeit this state has only a finite life time
$\tau_{\rm lt}$ and there is a diffusional time scale $\tau_D \gg
\tau_{\rm lt}$ where the ergodicity of the system is restored via
structural rearrangements of the particles.  Below $T_c$, the
decrease of the localization length $\xi$ with decreasing temperature
is accompanied by a rapid increase of $\tau_D$ and therefore also
with an increase of the life time $\tau_{\rm lt}$ of the amorphous
solid state, such that at sufficiently low temperatures below $T_c$,
the life time $\tau_{\rm lt}$ may reach macroscopic time scales.

One may expect that the response of a supercooled liquid to an
external mechanical load such as a shear field is qualitatively
different far below $T_c$ from the response above and around $T_c$.
This is due to the solid-like behavior over a large time scale
$\tau_{\rm lt}$ in the former case.  A system in an ideal amorphous
solid state (i.e.~with $\tau_{\rm lt}=\infty$) is associated with
a broken continuous translation symmetry which implies its rigidity
and the presence of long-range density correlations \cite{szamel2011}
as well as a far-field decay of frozen-in stress fluctuations
\cite{maier2017}.  When shearing a three-dimensional ideal amorphous
solid with a constant strain rate $\dot{\gamma}$ in a planar Couette
flow geometry, in the steady state, a flowing fluid state with a
constant shear stress $\sigma_{\rm ss}$ is obtained.  In the limit
$\dot{\gamma}\to 0$, the stress $\sigma_{\rm ss}$ is non-zero and
reaches the yield stress $\sigma_{\rm yield}$. Note that extensions
of MCT to glassforming liquids under shear have been proposed
\cite{fuchs2002, miyazaki2002, miyazaki2004, szamel2004}. In the
framework of the MCT by Fuchs and Cates \cite{fuchs2002}, a yield
stress is predicted for systems below $T_c$ \cite{amann2013,
amann2015}.

Thus, in an ideal amorphous solid, due to the broken translation
symmetry, the shear viscosity $\eta$ is infinitely large and one
does not obtain a Newtonian behavior with $\sigma_{\rm ss} = \eta
\dot{\gamma}$ in the limit $\dot{\gamma}\to 0$. However, this is
certainly different in a supercooled liquid far below $T_c$ that
is associated with a large but finite value of the time scale
$\tau_{\rm lt}$ on which it can be considered to be in an amorphous
solid state. In such a system, one expects on the one hand a Newtonian
behavior for $\dot{\gamma}^{-1}>\tau_{D}$ and on the other hand a
solid-like response for shear rates with $\dot{\gamma}^{-1}<\tau_{\rm
lt}<\tau_{D}$.  In the latter case, shear rates have to be sufficiently
small such that the resulting steady-state stress $\sigma_{\rm ss}$
is only slightly larger than an apparent yield stress that can be
obtained via extrapolation to the limit $\dot{\gamma}\to 0$ (see
below).

In this work, the latter regime is studied for a model glassformer
using non-equilibrium molecular dynamics (NEMD) computer simulation.
The model under consideration is a polydisperse soft-sphere system
that has been recently proposed by Ninarello {\it et
al.}~\cite{ninarello2017}.  It allows the application of the swap
Monte Carlo technique \cite{grigera2001} in combination with MD
simulation from which we obtain equilibrated samples far below
$T_c$, that we use as starting configurations for NEMD simulations
under shear. At sufficiently low shear rates, the simulations of
the sheared samples far below $T_c$ show features that, in computer
simulations, have been encountered so far only for out-of-equilibrium
glass states at very low or zero temperature.  In particular, we
observe the occurrence of brittle yielding \cite{schuh2007}, as
manifested by a sharp stress drop in the stress-strain relation at
a strain of the order of 0.1 \cite{ozawa2018, popovic2018, barlow2020}.
Thereby, we demonstrate that, for an appropriate choice of the shear
rate and temperature $T<T_c$, brittle yielding and shear banding
can be seen in a supercooled liquid state, provided that this state
exhibits transient elasticity over a significant time scale $\tau_{\rm
lt}$.

Our investigations are complementary to a recent study by Ozawa
{\it et al.}~\cite{ozawa2018} where, for the same model glassformer,
first fully equilibrated samples at different initial temperatures
$T_{\rm ini}$ above, around and far below $T_c$ were generated,
followed by a quench to zero temperature and subsequent shear
simulations using the athermal quasi-static shear (aqs) protocol.
As we shall see below, our findings are similar to those of Ozawa
{\it et al.}~when comparing the stress-strain relation of our shear
simulations at a given temperature $T$ and finite shear rate with
their aqs calculations for the corresponding temperature $T_{\rm
ini}=T$.  As in our case, they observe brittle yielding for
``well-annealed'' samples at $T_{\rm ini} \ll T_c$ while for
temperatures $T_{\rm ini}$ around and above $T_c$ a more ductile
response is seen.  The similar response in the aqs calculations and
our shear simulations is remarkable, keeping in mind that, in our
simulations, we shear supercooled liquids at a finite shear rate.
In the limit $\dot{\gamma}\to 0$, i.e.~in the ``quasi-static''
limit, these supercooled liquid states always show the ductile
mechanical response of a Newtonian liquid. This is also true for
temperatures below $T_c$ where elasticity has to be considered as
a transient phenomenon, albeit over a very long time scale $\tau_{\rm
lt}$ for temperatures far below $T_c$.  The fact that the aqs
simulations do not show a Newtonian response for initial temperatures
$T_{\rm ini}<T_c$ indicates that for well-annealed samples processes
that would lead to a Newtonian response are suppressed in the
framework of the aqs scheme and one obtains the response of a solid
with a finite yield stress.

The occurrence of brittle yielding is associated with the formation
of shear bands.  Shear banding is a ubiquitous phenomenon in glasses
under mechanical load \cite{schuh2007, ozawa2018, besseling2010,
divoux2010, chikkadi2011, divoux2016, maass2015, bokeloh2011,
binkowski2016, hubek2020, varnik2003, bailey2006, shi2006, shi2007,
ritter2011, sopu2011, chaudhuri2012, dasgupta2012, dasgupta2013,
albe2013, shiva2016_2, golkia2020, singh2020, parmar2019}. Especially
in metallic glasses, shear bands lead to inhomogeneities in the
microstructure and can cause a catastrophic failure of the material
\cite{schuh2007, maass2015, hubek2020}. In aqs simulations of a
glassforming binary Lennard-Jones mixture, Parmar {\it et
al.}~\cite{parmar2019} have demonstrated that shear-banded states
can be stabilized by applying oscillatory shear with an appropriate
strain amplitude, thereby obtaining states where a fluidized band
coexists with a stress-released amorphous solid.  This indicates
that at a given strain above the yield strain, shear-banded states
minimize the energy of the system.

Unlike previous studies, in this work, we observe brittle yielding
and shear banding in transient amorphous solids under equilibrium
conditions.  We find two types of shear-banded states right after
the yielding transition, namely states with horizontal and states
with vertical shear bands.  The formation of both types of shear
bands is an efficient way of releasing stresses, i.e.~the magnitude
of the stress drops is similar in both cases. However, in the case
of the vertical bands, the stress shows an increase with strain up
to a second maximum and a second, albeit smaller, stress drop which
is associated with the formation of a horizontal shear band in
addition to the vertical one. The formation of shear bands is also
associated with a drop of the potential energy such that, after the
drop, the potential energy is monotonously increasing towards the
steady state value. Recently, the occurrence of horizontal and
vertical shear bands has been also observed in sheared low-temperature
glass states of a binary Lennard-Jones mixture \cite{golkia2020};
however, in the present study, we find these features in equilibrated
systems.

The rest of the paper is organized as follows: In the next section
(Sec.~\ref{sec2}), the details of the model potential, the simulation
techniques, and the simulation protocols are reported.  Section 3
presents results on the equilibrium dynamics of supercooled liquids,
focussing on the change of the dynamics around the MCT critical
temperature. Section 4 is devoted to the analysis of supercooled
liquids under shear. Here, we address the question under which
conditions brittle yielding and shear banding occur. Finally, Sec.~5
summarizes the results and draws conclusions.


%
\section{Model and details of the simulation}
\label{sec2}
We consider a model of polydisperse non-additive soft spheres that
has been recently proposed by Ninarello {\it et al.}~\cite{ninarello2017}.
In this model, interactions between particles are pair-wise additive.
To each particle $i$, a diameter $\sigma_i$ is assigned according
to a probability distribution $P(\sigma) = A \sigma^{-3}$
with $A=2/(\sigma_{\rm min}^{-2}-\sigma_{\rm max}^{-2})$. We have
chosen $\sigma_{\rm min}=0.725\,\bar{\sigma}$ and $\sigma_{\rm
max}=\sigma_{\rm min}/(2\sigma_{\rm min}-1)\approx 1.611\,\bar{\sigma}$.
This choice of $\sigma_{\rm min}$ and $\sigma_{\rm max}$ provides
that the first moment of $P(\sigma)$ is equal to $\bar{\sigma)}$;
$\bar{\sigma}=1.0$ is used as the length unit in the following. The
interactions between pairs of particles depend on the variable
$x_{ij}=r_{ij}/\sigma_{ij}$ where $r_{ij} = | \vec{r}_i - \vec{r}_j
|$ is the distance between particle $i$ at position $\vec{r}_i$ and
particle $j$ at position $\vec{r}_j$ and $\sigma_{ij}=0.5 (\sigma_i
+ \sigma_j) (1- 0.2 |\sigma_i - \sigma_j|)$ introduces the
non-additivity of the particle diameters.  Note that the non-additivity
is essential to avoid any crystallization when the swap Monte
Carlo method is applied (see below).

The interaction potential between a pair of particles is defined by
\begin{equation}
  V(x) = 
   \begin{cases}
    V_0 (x^{-12} + c_0 + c_2 x^2 + c_4 x^4) & 
          \mathrm{for} \quad x < x_c\\
    0 & \mathrm{for} \quad x \geq x_c \, ,
   \end{cases}
\end{equation}
where the cut-off $x_c = 1.25$ is chosen. The terms with the
parameters $c_0 = -28/x_c^{12}$, $c_2 = 48/x_c^{14}$, $c_4 = -
21/x_c^{16}$ ensure the smoothness of the function $V(x)$ at $x=x_c$.
The parameter $V_0=1.0$ sets the unit of energy in the following.

The simulations at constant particle number $N$, constant volume
$V$, and constant temperature $T$ are performed with the LAMMPS
package \cite{plimpton1995}.  The number density is fixed at $\varrho
= N/V = 1.0$. The masses of the particles are set to $m=1.0$. In
the molecular dynamics (MD) simulations, Newton's equations of motion
are integrated by the velocity Verlet algorithm \cite{allenbook},
using a time step of $\delta t = 0.01\,\tau_{\rm MD}$ (with $\tau_{\rm
MD}=(\bar{\sigma}^2 m/V_0)^{1/2}$).  The temperature is kept fixed
by a DPD thermostat \cite{allenbook, soddemann2003}, using 
a similar implementation as in Ref.~\cite{golkia2020} with the
friction coefficient $\zeta=1.0$ and the weight function $\omega(r)= 1.0$
for $r\le 1.3 x_c$ and $\omega(r) = 0$ otherwise [cf.~Eqs.~(2)-(6) in 
Ref.~\cite{golkia2020}].
%
%
The DPD thermostat 
locally conserves the momentum
and is Galilean invariant. This is especially advantageous for the
non-equilibrium MD simulations under shear, because the Galilean-invariant
thermostat does not introduce any bias with respect to the direction
of the velocity flow. 

To obtain fully equilibrated samples at very low temperatures, a
combination of MD simulation and the swap-Monte-Carlo (SMC) technique
\cite{grigera2001} is used.  In a ``trial SMC move'', one randomly
selects a pair of particles and exchanges their diameters. Then,
this move is accepted or rejected according to a Metropolis criterion
\cite{allenbook}. In our hybrid scheme, every 25 MD steps $N$ trial
SMC moves are performed. In the considered temperature range, $0.01 \le
T \le 0.3$, the acceptance rate for the SMC moves varies between
10 and 22\% (with a decreasing acceptance rate with decreasing
temperature). The longest equilibration runs with the hybrid MD-SMC
method were over $10^7$ time steps which allowed to fully equilibrate
samples with $N=1372$, 2048, 6000, and 10000 particles at the
temperature $T=0.06$, corresponding to the glass transition temperature
$T_g$ in our study.

Non-equilibrium MD simulations are employed to shear the samples
in a planar Couette flow geometry. The shear is imposed via
Lees-Edwards boundary conditions \cite{lees1972} along the $xz$
plane in the direction of $x$. For the simulations under shear, we
have integrated the equations of motion with the time step $\delta
t = 0.001\,\tau_{\rm MD}$. Most of the data shown below correspond
to the temperatures $T=0.15$, 0.11, 0.09, 0.07, and 0.06 for a
system of $N=10000$ particles. At each temperature, 30 runs were
performed, starting from statistically independent samples that
were fully equilibrated via the MD-SMC method. The considered shear
rates range from $\dot{\gamma} = 10^{-6}$ to $\dot{\gamma} = 10^{-3}$.
For the calculation of the stress-strain relations, we have performed
a running average over strain windows of width $\delta \gamma =
10^{-4}$.

\section{From liquid to amorphous solid: Equilibrium dynamics}
The dynamics of supercooled liquids is associated with the cage
effect. At sufficiently low temperatures, the particles are trapped
in cages formed by the surrounding particles and the breaking of
cages requires collective particle rearrangements that slow down
with decreasing temperature.  As we shall see below, around the
critical temperature of mode coupling theory (MCT), $T_c$, the
system gradually transforms from a liquid-like state to a state
that can be characterized as an amorphous solid. This transition
is due to the localization of the particles in their cages and, as
we shall see in the next section, the response to an external shear
changes drastically from the liquid-like state above $T_c$ to the
amorphous solid well below $T_c$, especially with respect to the
yielding behavior. In this section, we first present the ``equation
of state'' of our system, i.e.~the temperature dependence of the
potential energy per particle, and then study the one-particle
dynamics in terms of the mean-squared displacement (MSD) of a tagged
particle. From the MSD, a localization length is determined that
indicates the transition from liquid to solid-like behavior around
$T_c$. Furthermore, we estimate the life time $\tau_{\rm lt}$ of 
the amorphous solid as a function of temperature. We have
computed the equation of state from fully equilibrated configurations
that we have obtained via hybrid MD-SMC simulations at constant temperature.
For the calculation of the MSD, we have used such fully equilibrated 
samples as starting configurations for microcanonical runs where
we have switched off the SMC and the coupling to the thermostat.

\begin{figure}[htb!]
\centering
\includegraphics[width=7.5cm]{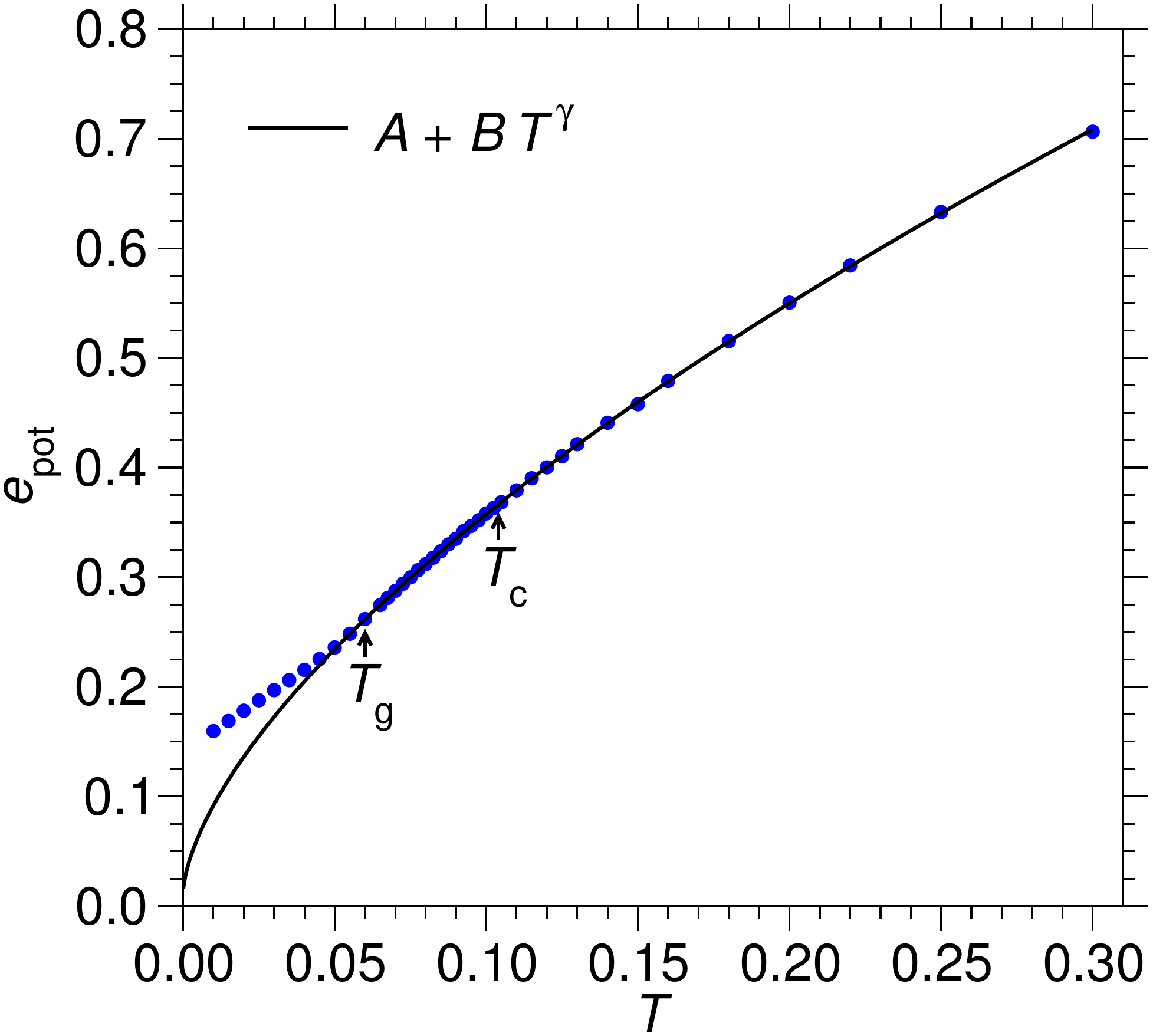}
\caption{Potential energy per particle, $e_{\rm pot}$, as a function
of temperature $T$. The solid line is a fit with the function $f(T)
= A + B T^{\gamma}$ with $A= 0.0120017$, $B=1.4999$, and $\gamma =
0.637135$. Indicated are the locations of the critical
mode coupling temperature, $T_c = 0.104$, and the glass transition
temperature, $T_g = 0.06$. \label{fig1}}
\end{figure}
Figure \ref{fig1} shows the potential energy per particle, $e_{\rm
pot}$, as a function of temperature. In this plot, the critical MCT
temperature at $T_c = 0.104$ as well as the glass transition
temperature at $T_g = 0.06$ are indicated.  The MCT temperature
$T_c$ was determined from fits to dynamic quantities such as the
mean-square displacement (see below). Below $T_g$, the hybrid
MD-SMC runs on the time scale of $10^5\,\tau_{\rm MD}$ are no longer
sufficient to fully equilibrate the system. The data for $T\ge
T_g$ can be well described by the function (solid line in 
Fig.~\ref{fig1})
\begin{equation}
f(T) = A + B T^{\gamma} 
\label{eq1}
\end{equation}
with $A$, $B$, and $\gamma$ being fit parameters. While the density
functional theory of Rosenfeld and Tarazona \cite{rosenfeld1998}
predicts the exponent $\gamma = 0.6$ for simple high-density
soft-sphere fluids, we find the exponent $\gamma\approx 0.64$ which
is very close to this prediction. Note that Eq.~(\ref{eq1}) with a
value of $\gamma$ around 0.6 also provides a good approximation for
other glassforming liquids with a $1/r^n$-type interactions at low
temperature (for a detailed discussion see Ref.~\cite{ingebrigtsen2013}).

\begin{figure}
\centering
\includegraphics*[width=7.25cm]{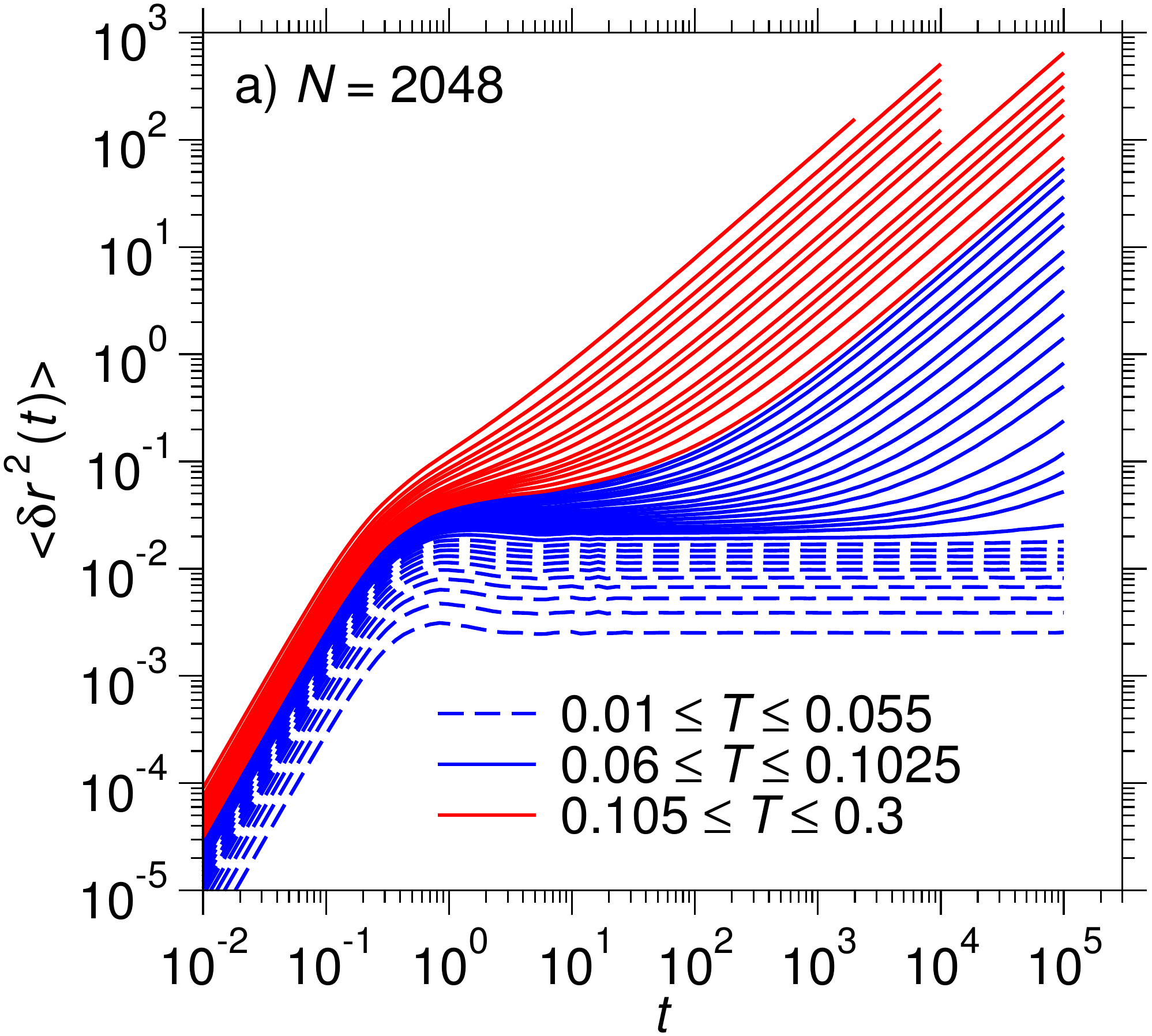}
\includegraphics*[width=7.25cm]{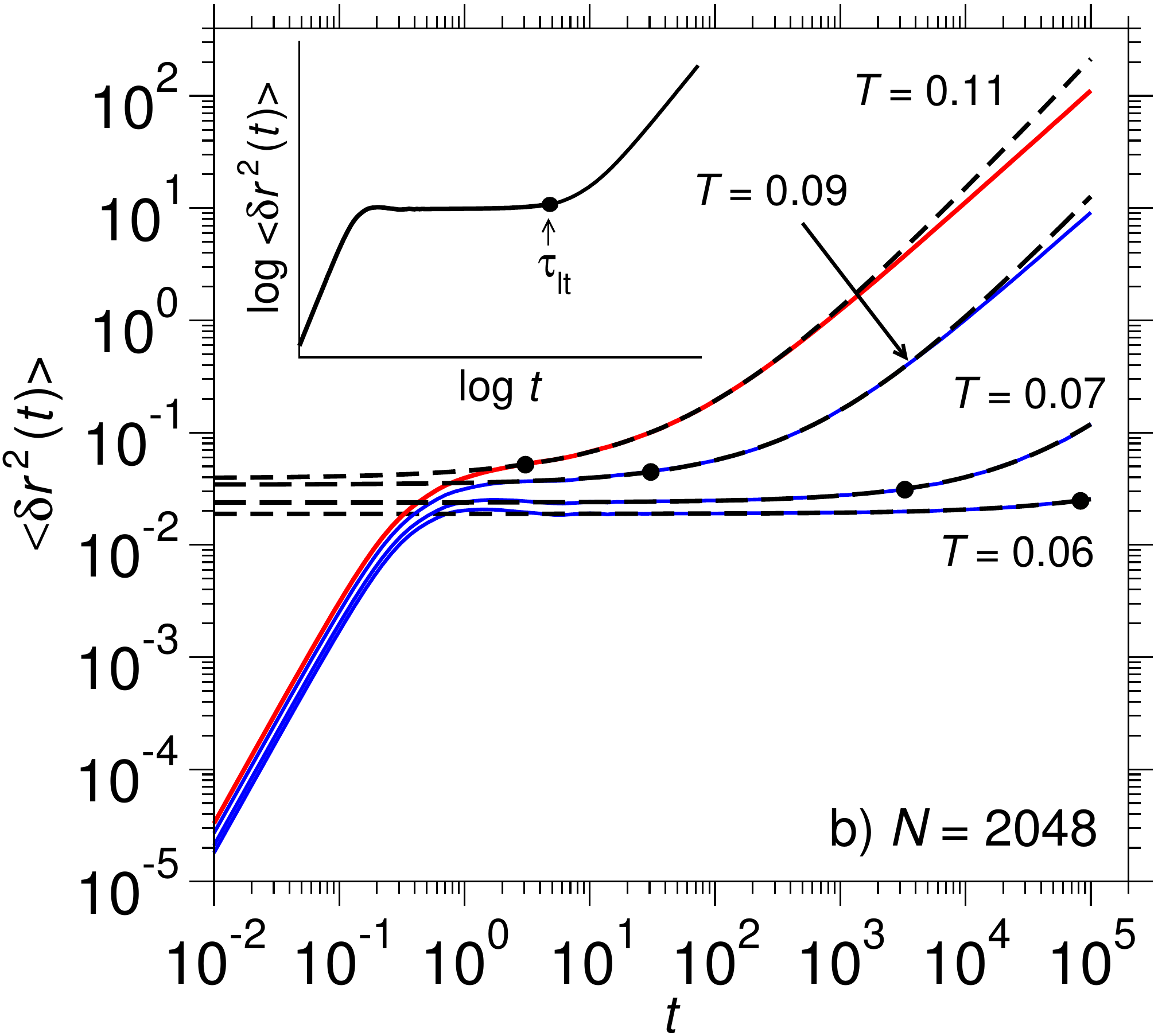}
\includegraphics*[width=7.25cm]{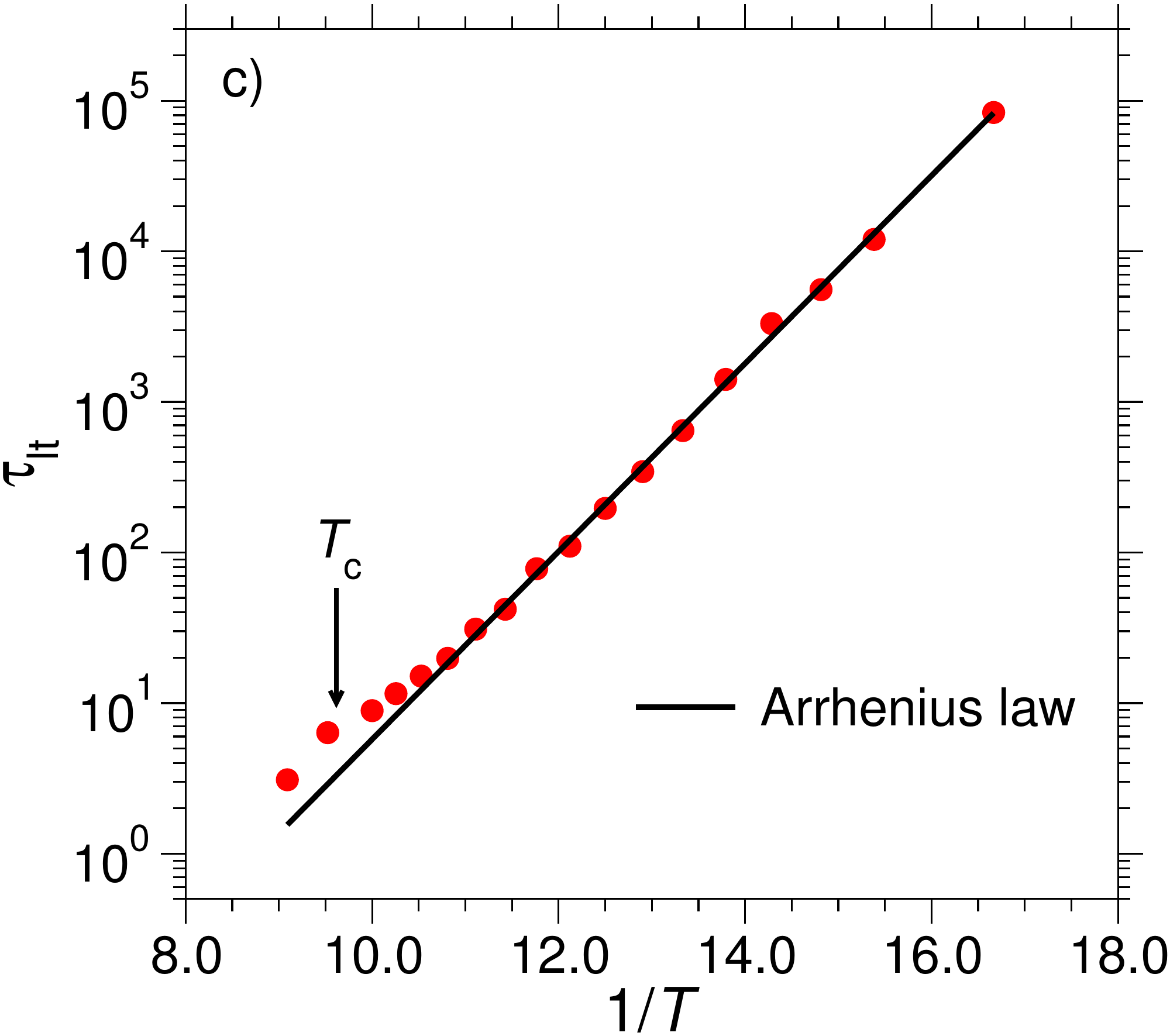}
\caption{a) MSD as a function of time for temperatures $T < T_g$
(dashed blue lines), $T_g \le T < T_c$ (solid blue lines), and $T
> T_c$ (red solid lines). b) MSDs at $T=0.11$, $T=0.09$, $T=0.07$,
and $T=0.06$. The dashed lines are fits to Eq.~(\ref{eq_phi}) and
the filled circles mark the location of the life time $\tau_{\rm
lt}$ for the different temperatures (see text). The inset is a
schematic plot of the MSD that illustrates the definition of
$\tau_{\rm lt}$. c) Life time $\tau_{\rm lt}$ as a function of 
inverse temperature. The solid line is a fit with an Arrhenius 
law (see text). The MSDs in a) and b) correspond to systems with
$N=2048$ particles.  \label{fig2}}
\end{figure}
Now we come to the one-particle dynamics of the system and investigate
the MSD of a tagged particle, defined by
\begin{equation}
\left\langle \delta r^2(t) \right\rangle =
\frac{1}{N} \sum_{i=1}^N \left\langle 
|\vec{r}_i(t) - \vec{r}_i(0)|^2 \right\rangle
\label{eq_msd}
\end{equation}
with $\vec{r}_i(t)$ the position of particle $i$ at time $t$. The
brackets $\langle \dots \rangle$ represent an ensemble as well as
a time average over the different samples. Note, however, that for
states below $T_g$ we have only applied an ensemble average. The
MSDs are calculated from microcanonical MD simulation for a system
of $N=2048$ particles, using as initial configurations 60 independent
samples from the MD-SMC simulations.

In Fig.~\ref{fig2}a, MSDs are plotted double-logarithmically for
different temperatures. Here, we have marked the different temperature
regimes.  The red solid lines correspond to temperatures above $T_c$
at $T=0.105$, 0.11, 0.115, 0.12, 0.125, 0.13, 0.14, 0.15, 0.16,
0.18, 0.20, 0.22, 0.25, and 0.3. At the highest temperature, $T=0.3$,
the MSD displays a ballistic regime $\propto t^2$ at very short
times, an emerging shoulder at intermediate times, and a diffusive
regime $\propto t$ in the long-time limit. With decreasing temperature,
the diffusive regime shifts to longer times and the intermediate
time regime evolves into a plateau. The blue solid lines show the
MSDs for temperatures $T_g < T < T_c$ at $T=0.06$, 0.065, 0.0675,
0.07, 0.075, 0.0775, 0.08, 0.0825, 0.085, 0.0875, 0.09, 0.0925,
0.095, 0.0975, 0.10, and 0.1025. Here, the initial configurations
are fully equilibrated samples from the MD-SMC simulations. However,
the microcanonical MD runs over a time scale of $10^5\,\tau_{\rm
MD}$ are not long enough to reach a diffusive regime far below
$T_c$.  So at $T=0.06$, we hardly see deviations from the plateau
at long times. The MSDs below $T_g$ in Fig.~\ref{fig2}a (blue dashed
lines) correspond to the temperatures $T=0.01$, 0.015, 0.02, 0.025,
0.03, 0.035, 0.04, 0.045, 0.05, and 0.055. Here, the MSDs display
a plateau for $1 \le t \le 10^5$, the height of which decreases
with decreasing temperature. Note that the small overshoot in the
low-temperature MSDs around $t\approx 0.8$ is associated with the
microscopic dynamics \cite{horbach1996, horbach2001}.  This feature
disappears for larger system sizes (e.g.~for our model it cannot
be seen anymore for systems with $N=10000$ particles).

The emergence of a shoulder that evolves into a plateau at low
temperature manifests the caging of the particles. MCT provides
detailed predictions about the behavior of the MSD around the plateau
(as well as corresponding predictions for the plateau-like regions
in intermediate scattering functions \cite{goetze2009}).
One of them describes the initial increase of the MSD from the 
plateau and is given by \cite{goetze2009}  
\begin{equation}
\phi(t) = \delta r^2_{\rm plateau} + h t^b + h_2 t^{2b} \, .
\label{eq_phi}
\end{equation}
This equation corresponds to a von Schweidler law, extended by a
correction term $\propto t^{2b}$. $\delta r^2_{\rm plateau}$
quantifies the height of the (emerging) plateau in the MSD, $h$ and
$h_2$ are temperature-dependent amplitudes, and the exponent $b$
is expected to be universal for a given system (but it may vary for
different systems in the range $0< b \le 1$). Figure \ref{fig2}b
shows the MSDs at $T=0.11$, $T=0.09$, $T=0.07$, and $T=0.06$ together
with fits to Eq.~(\ref{eq_phi}). These fits and also the fits to
the MSDs at the other temperatures were performed with the constant
exponent value $b=0.59$. Note, however, that the values for $\delta
r^2_{\rm plateau}$, as obtained from the fit to Eq.~(\ref{eq_phi}),
are not very sensitive with respect to the choice of the exponent
$b$.

Using the fits to Eq.~(\ref{eq_phi}), we can now introduce a
definition of the life time $\tau_{\rm lt}$ of the transient amorphous
solid state for the different temperatures. To this end, we define
$\tau_{\rm lt}$ as the time for which $\left\langle \delta r^2(\tau_{\rm
lt}) \right\rangle/ \delta r^2_{\rm plateau} = 1.3$ (see the inset
of Fig.~\ref{fig2}b for an illustration of this definition). The
locations of $\tau_{\rm lt}$ for the MSDs in Fig.~\ref{fig2}b are
marked as filled circles.

Figure \ref{fig2}c shows the logarithm of the time scale $\tau_{\rm
lt}$ as a function of inverse temperature. For $T\lesssim 0.09$,
the data can be well fitted by an Arrhenius law $f(T)=\tau_0 \,
\exp\left( E_{\rm A}/T \right)$, which is represented by the bold
solid line in the figure.  The values of the fit parameters are
$\tau_0 = 3.3\times 10^{-6}$ and $E_{\rm A} = 1.43641$.  Here, the
energy $E_{\rm A}$ can be interpreted as an activation energy.  The
application of the Arrhenius law and thus the interpretation of a
kinetic process as an activated one are only sensible if the ratio
of the activation energy to the thermal energy, $E_{\rm A}/T$, is
much larger than unity \cite{riskenbook}. In our case, this ratio
varies between about 16 at $T=0.09$ and about 24 at $T=0.06$ which
is consistent with the condition $E_{\rm A}/T \gg 1$.  At temperatures
$T\gtrsim T_c$, we observe significant deviations from the Arrhenius
behavior and $\tau_{\rm lt}$ is close to the microscopic time scale
$\tau_{\rm MD}$.  From the temperature dependence of $\tau_{\rm
lt}$ we can conclude that around $T_c$ there is a gradual crossover
towards an activated dynamics with decreasing temperature.

\begin{figure}
\centering
\includegraphics*[width=7.5cm]{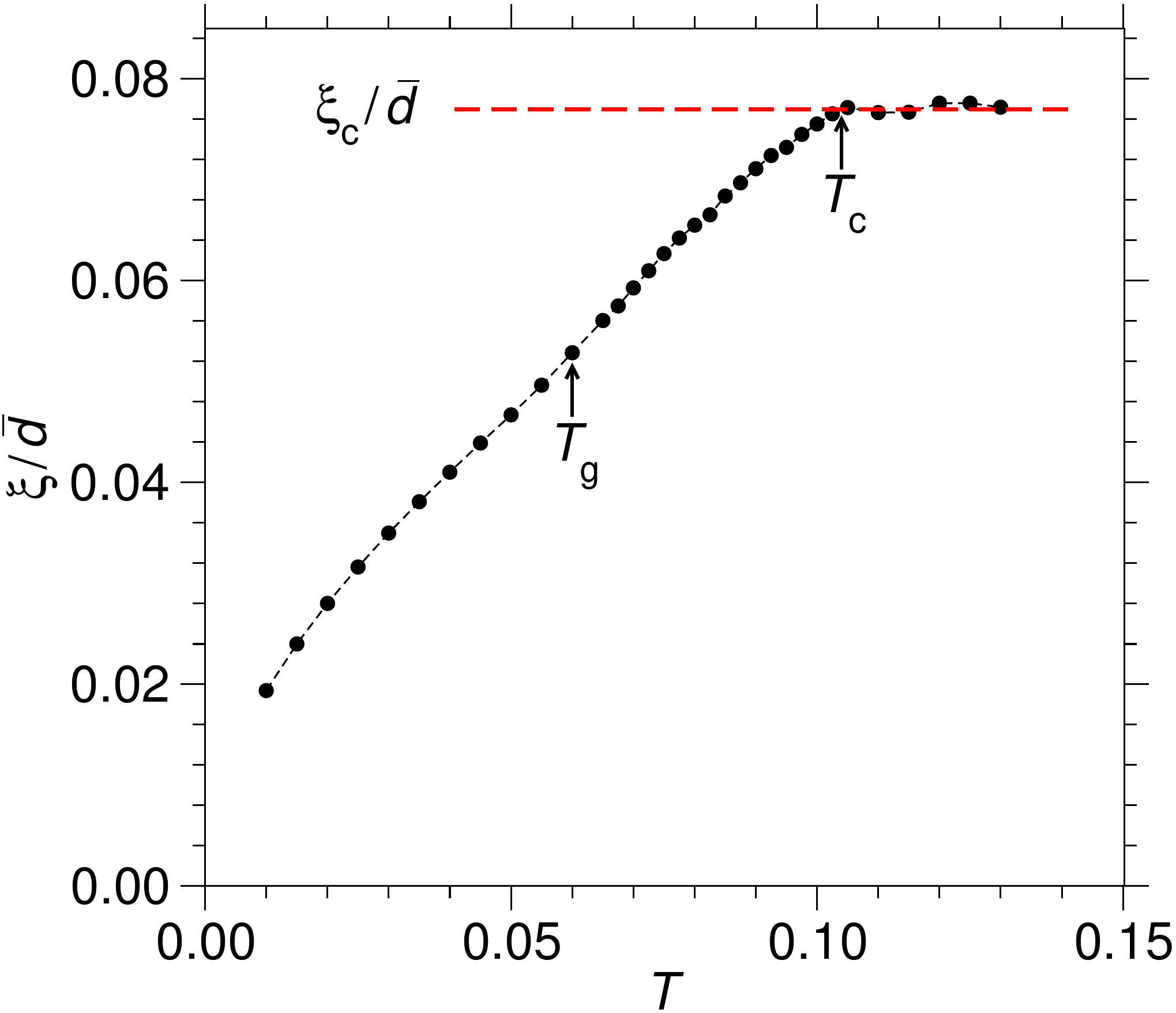}
\caption{Localization length divided by the mean nearest-neighbor 
distance, $\xi/\bar{d}$, as a function of temperature. The dashed
red line marks the critical value of the reduced localization length,
$\xi_c/\bar{d} \approx 0.077$.  
\label{fig3}}
\end{figure}
In the framework of the Gaussian approximation \cite{hansenbook,
thorneywork2016, fuchs1998}, one can relate $\delta r^2_{\rm plateau}$
to a localization length $\xi$ as
\begin{equation}
 \xi^2 = \frac{1}{6} \delta r^2_{\rm plateau} \, .
\label{eq_xi2}
\end{equation}
Figure \ref{fig3} shows the temperature dependence of $\xi$, scaled
with the average nearest-neighbor distance $\bar{d} \approx 1.07$
(we have estimated $\bar{d}$ from the location of the first peak
of the radial distribution function at $T=0.06$).  At $T=0.01$, i.e.~far below
$T_g$, the reduced localization length is $\xi/\bar{d}\approx 0.02$.
It increases with increasing temperature.  At $T_g$, $\xi/\bar{d}$
slightly changes slope and then increases roughly linearly up to
$T_c$ where it reaches the constant $\xi_c/\bar{d}\approx 0.077$.
The critical value, $\xi_c$, of the localization length marks the
stability limit of the amorphous solid, i.e.~for $T>T_c$ the system
is in a liquid state. In analogy to crystalline solids, the critical
value $\xi_c/\bar{d}$ can be interpreted as a Lindemann criterion
for the stability of an amorphous solid \cite{goetze2009}.  Note
that Fuchs {\it et al.}~\cite{fuchs1998} have obtained
$\xi_c/\bar{d}\approx 0.0746$ in a calculation for a hard sphere
system in the framework of MCT, thus a value that is very close to
our finding.

The behavior of both $\tau_{\rm lt}$ and $\xi$ indicate a gradual
change of the dynamics around $T_c$.  Below $T_c$, the localization
of particles in their cages, as quantified by $\xi_c/\bar{d}$, is
below the stability limit, given by $\xi_c/\bar{d}\approx 0.077$.
As a consequence, there is the emergence of transient amorphous
solid state for $T<T_c$, the life time $\tau_{\rm lt}$ of which
follows an Arrhenius law with an activation energy of about 1.44.
The gradual change from liquid-like to solid-like dynamics is also
associated with a qualitative change of the system's response to
an external shear. As we shall see in the next section, brittle
yielding and the formation of shear bands can be observed in the
supercooled liquid below $T_c$. These features are typical for the
response of low-temperature glasses to a mechanical load.  In the
following, we shall analyze the conditions for the occurrence of
brittle yielding and shear banding in deeply supercooled liquids.
An important parameter in this context is the time scale $\tau_{\rm
lt}$.  For example, for $T=0.06$, the life time $\tau_{\rm lt}$ is
of the order of $10^5$ (Fig.~\ref{fig2}c).  Therefore, for
$\dot{\gamma}\gtrsim 10^{-5}$, the product $\dot{\gamma}\tau_{\rm
lt}$ is lower equal unity and one may expect the shear response of
an amorphous solid.

\section{Supercooled liquids under shear}
Now we analyze the results for equilibrated supercooled liquids
under shear.  Our focus is on the temperature range $0.06 \le T \le
0.15$ to study the response to the external shear from liquid-like
states slightly above $T_c$ to the solid states far below $T_c$.
As we have seen in the previous section, the latter states can be
characterized via the localization length $\xi$ being significantly
lower than the critical value $\xi_c$.

\begin{figure}
\centering
\includegraphics*[width=7.5cm]{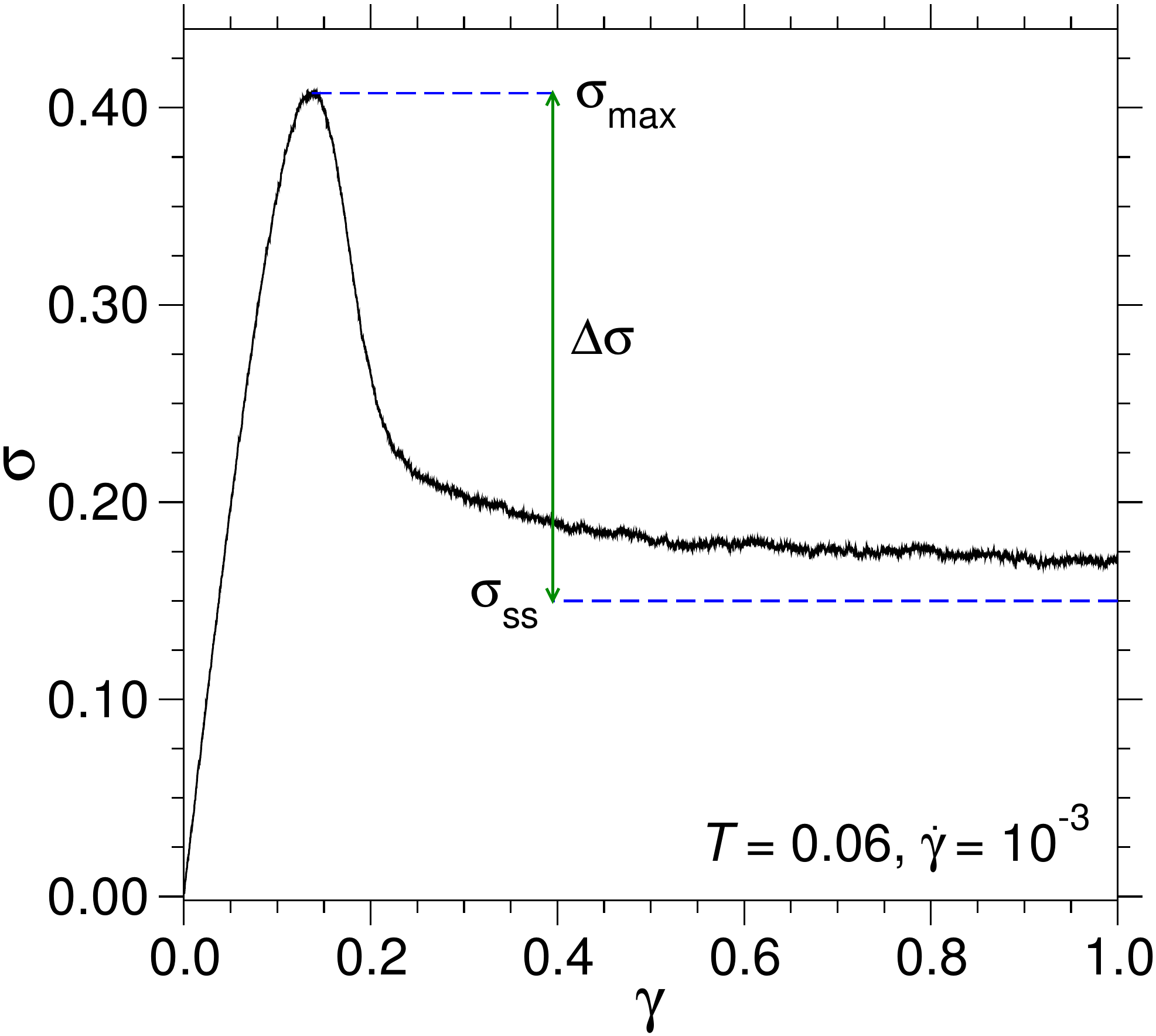}
\caption{Stress-strain relation for the temperature $T=0.06$ and
the shear rate $\dot{\gamma}=10^{-3}$. Indicated are the maximum
$\sigma_{\rm max}$, the steady-state stress $\sigma_{\rm ss}$, and
the definition of the stress drop $\Delta \sigma$. \label{fig4}}
\end{figure}
A typical stress-strain relation, indicating a non-Newtonian response
of the supercooled liquid, is shown in Fig.~\ref{fig4} for the
temperature $T=0.06$ and the shear rate $\dot{\gamma}=10^{-3}$.
While the strain is given by $\gamma = \dot{\gamma} t$, the stress
$\sigma$ was computed from the virial equation, as described in
Ref.~\cite{golkia2020}.  Different regimes can be identified in the
figure. First the stress increases almost linearly up to a maximum
value $\sigma_{\rm max}$ which is reached at a strain $\gamma \approx
0.135$ in this case.  The maximum in the stress marks the transition
from an elastic deformation of the ``solid'' to the onset of plastic
flow.  During the plastic deformation, the stress drops from
$\sigma_{\rm max}$ towards the steady-state stress $\sigma_{\rm
ss}$ which can be quantified by $\Delta \sigma = \sigma_{\rm max}
- \sigma_{\rm ss}$.  In the steady state, the system can be described
by a flowing homogeneous liquid.  Heterogeneous flow patterns are
observed between the onset of plastic flow and the steady state.
The morphology of these flow patterns, especially with respect to
the dependence on temperature and shear rate, shall be elucidated
in the following.

\begin{figure}
\centering
\includegraphics*[width=7.5cm]{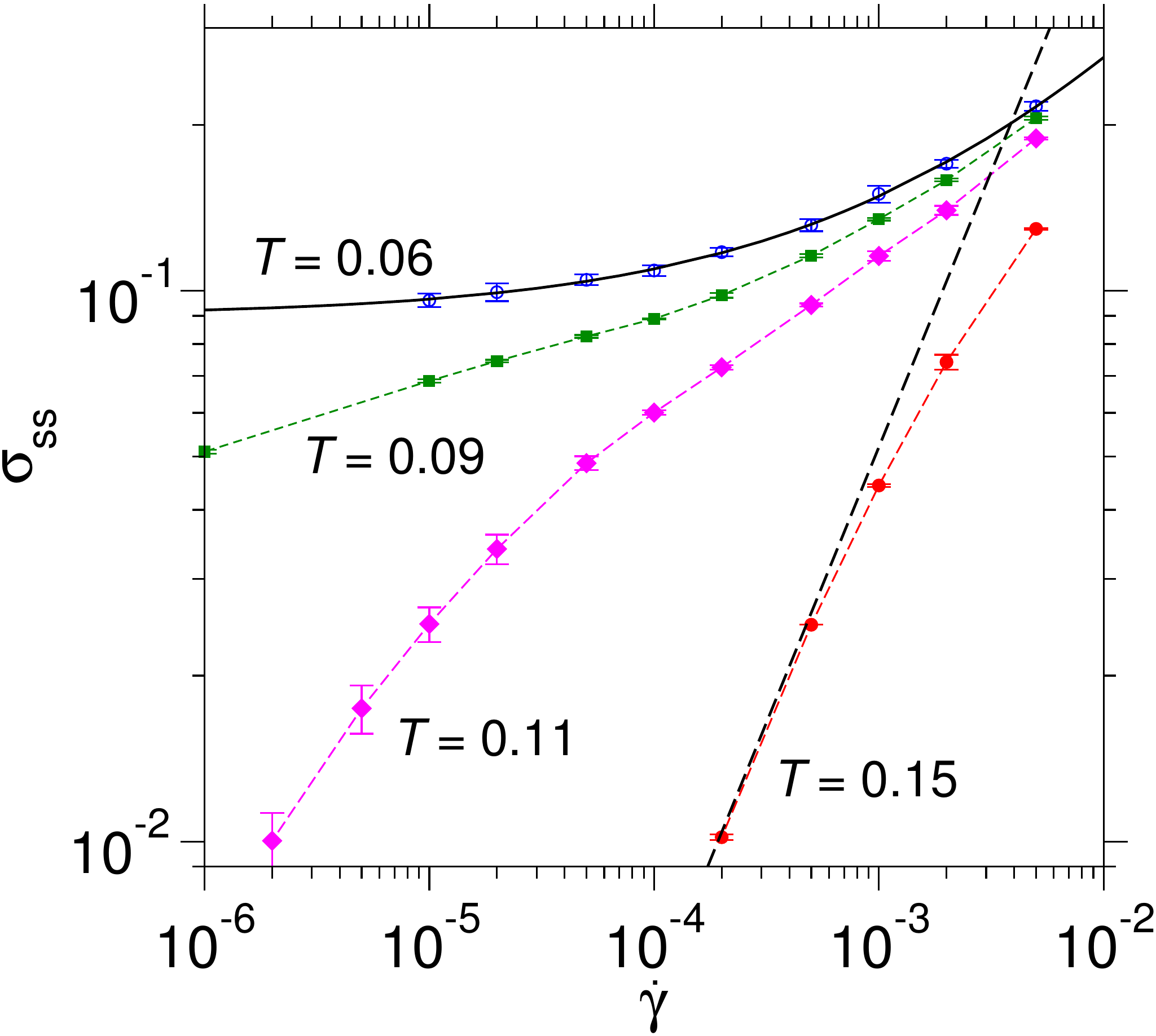}
\caption{Flow curves for the temperatures $T=0.06$, 0.09, 0.11, and
0.15. The solid line is a fit with a Herschel-Bulkley law to the
data for $T=0.06$ (see text). The dashed line indicates a linear
behavior, $\sigma_{\rm ss} \propto \dot{\gamma}$. \label{fig5}}
\end{figure}
In Fig.~\ref{fig5}, the steady-state stress $\sigma_{\rm ss}$ as a
function of the shear rate $\dot{\gamma}$ (i.e.~the flow curve) is
plotted double-logarithmically for different temperatures above and
below $T_c$. For sufficiently low shear rates, one expects that the
system behaves like a Newtonian fluid with a linear increase of the
stress as a function of the shear rate, $\sigma_{\rm ss} = \eta
\dot{\gamma}$ (with $\eta$ the shear viscosity).  At $T=0.15$, we
can still identify a Newtonian regime (dashed line), followed by
sublinear shear-thinning regime for $\dot{\gamma}>5\times 10^{-4}$.
At $T=0.11$, the Newtonian regime is not anymore in the window of
considered shear rates $\dot{\gamma}\ge 10^{-6}$. Here, we observe
an emerging plateau around $\dot{\gamma} = 10^{-4}$ that becomes
more pronounced at $T=0.09$, and eventually, at $T=0.06$, the data
can be well fitted by a Herschel-Bulkley law \cite{herschel1926}
(solid line), $\sigma_{\rm ss} = \sigma_{\rm yield} + A
\dot{\gamma}^\alpha$ with the yield stress $\sigma_{\rm yield}=
0.0900265$, the amplitude $A=1.57446$, and the exponent $\alpha=0.477008$.
Note that at $T=0.07$ the flow curve is very similar to that at
$T=0.06$.  So at the lowest considered temperatures where we are
able to obtain a fully equilibrated state, our system can be seen
as a yield stress material in equilibrium (although we also expect
at these temperatures the occurrence of a Newtonian regime at
extremely low shear rates).

\begin{figure}
\centering
\includegraphics*[width=7.5cm]{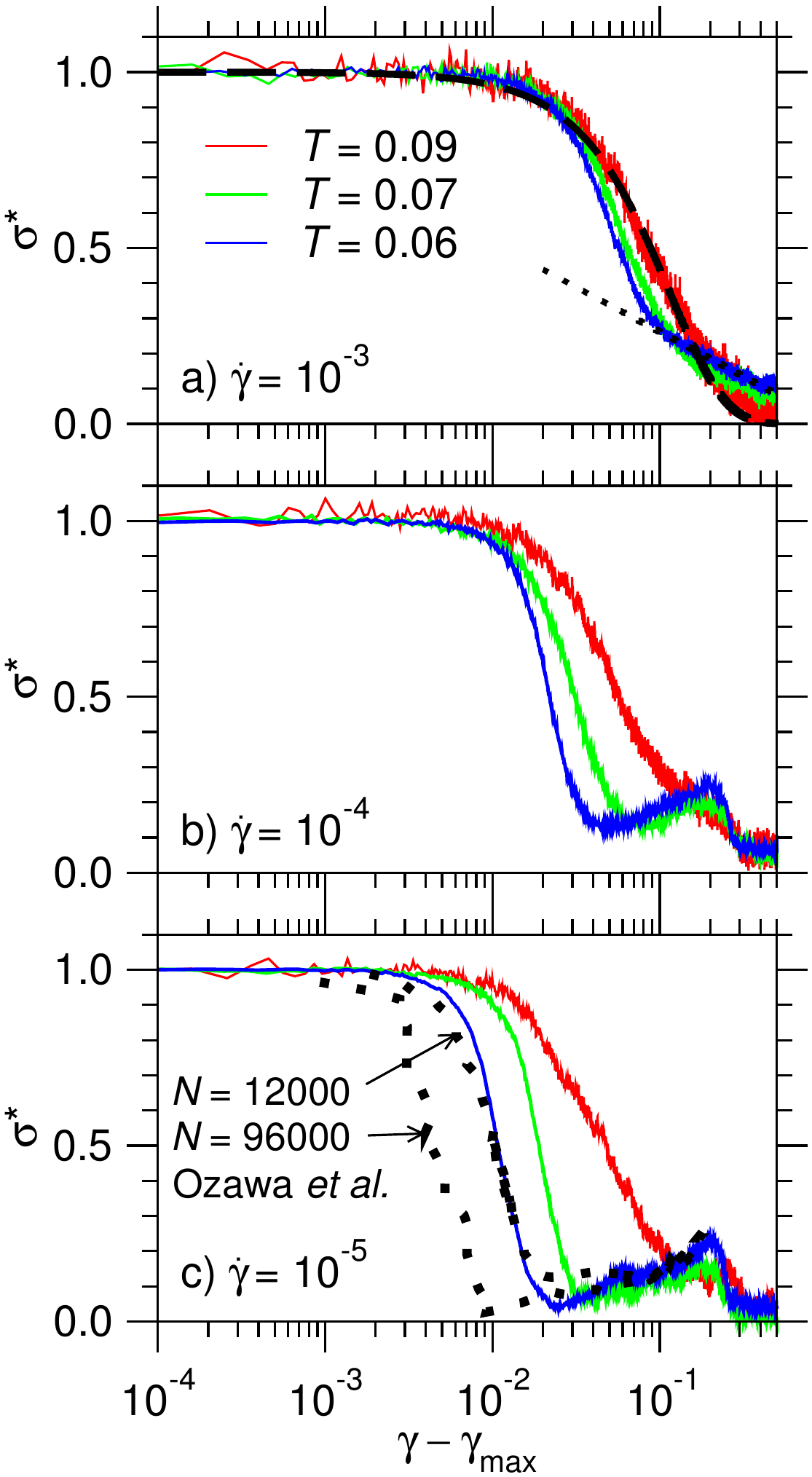}
\caption{Reduced stress $\sigma^\star$ as a function of $\gamma-\gamma_{\rm
max}$ at the different temperatures $T=0.06$, 0.07, and 0.09 for
the shear rates a) $\dot{\gamma} = 10^{-3}$, b) $\dot{\gamma} =
10^{-4}$, and c) $\dot{\gamma} = 10^{-5}$.  The data corresponds
to systems with $N=10000$ particles.  In c), the dotted lines
correspond to data for $N=12000$ and $N=96000$, as adapted from
Ref.~\cite{ozawa2018}.  The dashed and the dotted line in a) are
fits with with a compressed exponential and a logarithm, respectively
(see text). \label{fig6}}
\end{figure}
Having characterized the steady-state behavior of our system under
shear, we now investigate the relaxation of the stress from the
onset of plastic flow (marked by the maximum stress $\sigma_{\rm
max}$ at a given temperature) to the steady-state stress. To this
end, we define the reduced stress
\begin{equation}
\sigma^\star = 
\frac{\sigma - \sigma_{\rm ss}}{\sigma_{\rm max}-\sigma_{\rm ss}}
\label{eq_sigmastar}
\end{equation}
which is displayed in Fig.~\ref{fig6} for three shear rates and
three temperatures below $T_c$ as a function of $\gamma - \gamma_{\rm
max}$ (with $\gamma_{\rm max}$ the strain corresponding to $\sigma_{\rm
max}$).  At $\dot{\gamma} = 10^{-3}$ (Fig.~\ref{fig6}a), the decay
of $\sigma^\star$ for $T=0.09$ can be described by the compressed
exponential $\exp[- ((\gamma - \gamma_{\rm max})/\delta
\gamma^\star)^{a_{\rm ce}}]$ with $\delta \gamma^\star = 0.115946$
and $a_{\rm ce} = 1.33297$.  Also for the two lower temperatures
$T=0.06$ and 0.07, the reduced stress decays on a strain scale
$\delta \gamma^\star \approx 0.1$, but the functional form of its
decay changes around $\gamma - \gamma_{\rm max} \approx 0.08$ in
that the compressed-exponential-like decay is followed by a logarithmic
one $\propto \ln[ (\gamma - \gamma_{\rm max})/1.07353]$ (dotted
line in Fig.~\ref{fig6}a, fitted to the ``tail'' of the $T=0.06$
curve).  The difference in the decay of $\sigma^\star$ with respect
to temperature becomes more pronounced at the lower shear rates
$10^{-4}$ (Fig.~\ref{fig6}b) and $10^{-5}$ (Fig.~\ref{fig6}c). While
at $T=0.09$, the reduced stress still decays essentially with a
compressed exponential on the strain scale $\delta \gamma^\star
\approx 0.1$, at the two lower temperatures the initial decay is
significantly faster and $\sigma^\star$ exhibits a local maximum
around $\gamma - \gamma_{\rm max} \approx 0.2$.  The strain scale
of the initial decay decreases with decreasing temperature and shear
rate. At $T=0.06$ and $\dot{\gamma}=10^{-5}$, the reduced stress
$\sigma^\star$ decays on the strain scale $\delta \gamma^\star
\approx 0.01$.  

Also included in Fig.~\ref{fig6}c are data for $N=12000$ and
$N=96000$, as adapted from the simulation study of Ozawa {\it et
al.}~\cite{ozawa2018} using an aqs protocol. It is remarkable that
the reduced stress for $N=12000$ from Ozawa {\it et al.}~agrees
well with our data for a comparable system of $N=10000$ particles,
although we consider a system at a finite temperature as well as a
finite strain rate and, moreover, our system is in an equilibrated
supercooled liquid state (note, however, that $\sigma_{\rm max}$
is significantly larger in the athermal case).

\begin{figure}
\centering
\includegraphics*[width=7.5cm]{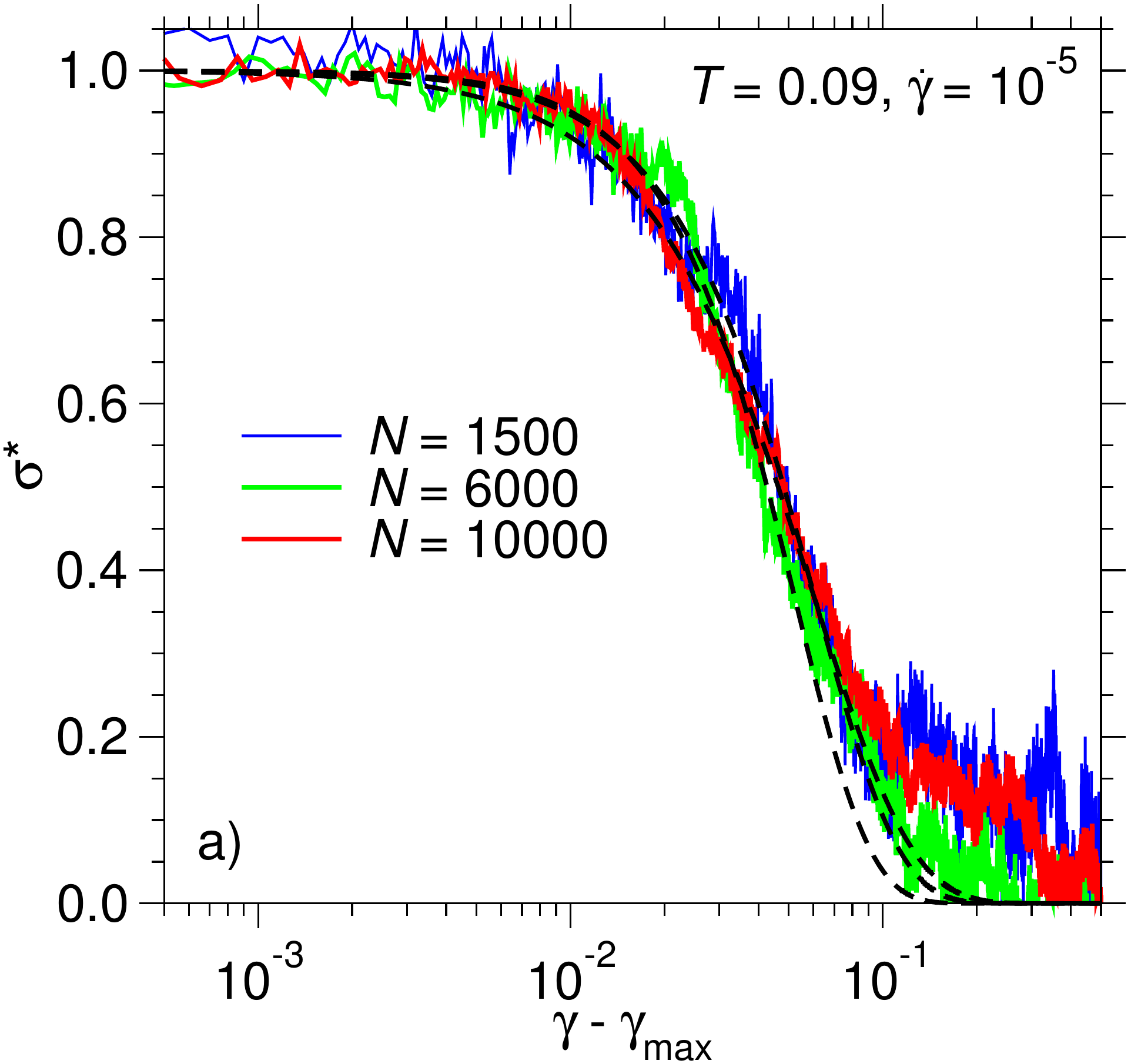}
\includegraphics*[width=7.5cm]{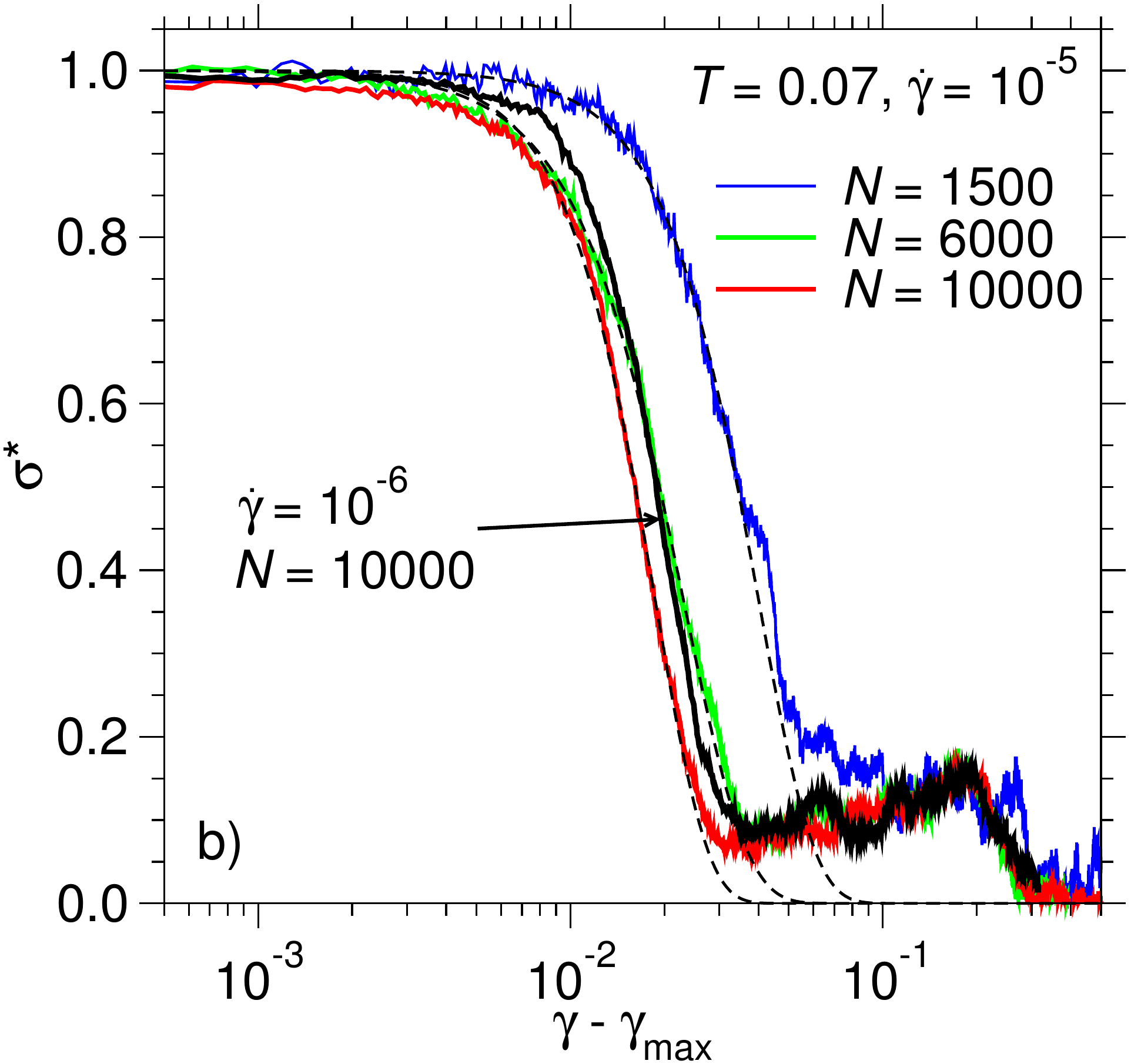}
\includegraphics*[width=7.5cm]{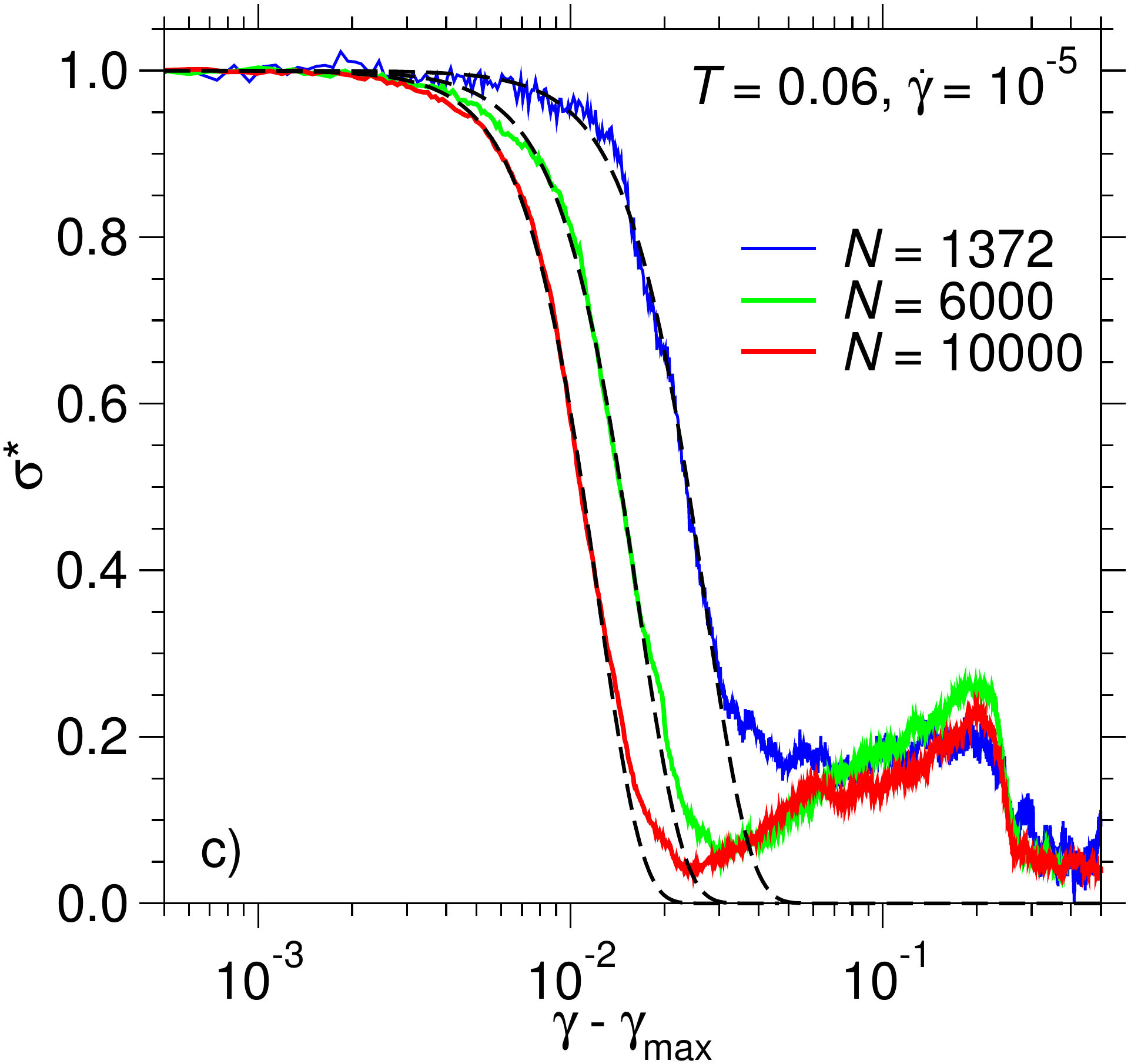}
\caption{Reduced stress $\sigma^\star$ as a function of $\gamma -
\gamma_{\rm max}$ at the shear rate $\dot{\gamma} = 10^{-5}$ and
temperatures a) $T=0.09$, b) $T=0.07$, and c) $T=0.06$ for different
system sizes, as indicated. The dashed lines are fits with compressed
exponentials (see text). Also included in b) is the reduced stress
for $\dot{\gamma}=10^{-6}$ and $N=10000$. \label{fig7}}
\end{figure}
It is tempting to interpret the rapid drop of the stress as a
first-order phase transition, as proposed by Ozawa {\it et
al.}~\cite{ozawa2018}.  The interpretation of the stress drop as a
phase transition would be appropriate in the limit of zero shear
rate, $\dot{\gamma}\to 0$. So we have to take this limit in some
sensible manner, keeping in mind that the expected true behavior
of the system in the zero shear-rate limit is that of a Newtonian
fluid for which $\sigma_{\rm ss}\propto \dot{\gamma}$ and the absence
of any stress drop in the stress-strain relation. However, the fluid
curves for $T\le 0.07$ suggest that the systems can be considered
as a yield stress fluid also at very low shear rates and one obtains
$\sigma_{\rm yield}$ by extrapolation via the Herschel-Bulkley law.
Below, we perform a similar extrapolation to obtain the initial
strain scale $\delta \gamma^\star$ with which the stress decays from
$\sigma_{\rm max}$ to $\sigma_{\rm ss}$ in the limit $\dot{\gamma}\to 0$.

The existence of a first-order transition in the thermodynamic limit
implies that it is rounded for finite systems and it becomes sharper
with increasing system size. Figure \ref{fig7} shows the decay of
the reduced stress $\sigma^\star$ for different system sizes at the
temperatures $T=0.09$, $T=0.07$, and $T=0.06$ in panels a), b), and
c), respectively. For all three temperatures, the shear rate is
$\dot{\gamma} = 10^{-5}$. While for $T=0.09$ there is almost no
dependence of $\sigma^\star$ on system size, for the two lower
temperatures the decay of $\sigma^\star$ becomes significantly
sharper with increasing system size. For $T=0.07$ (Fig.~\ref{fig7}b),
we have also included the reduced stress for the lower shear rate
$\dot{\gamma}=10^{-6}$ and $N=10000$ which exhibits a less rapid
decay than the corresponding result for $\dot{\gamma}=10^{-5}$.
This can be explained in terms of the life time $\tau_{\rm lt}$ in
relation to the shear rate $\dot{\gamma}$.  Above we have estimated
$\tau_{\rm lt}\approx 3300$ for $T=0.07$ (cf.~Fig.~\ref{fig2}c) and thus the time scale
$\dot{\gamma}^{-1}$ is much larger than $\tau_{\rm lt}$ for both
shear rates $10^{-5}$ and $10^{-6}$. The yielding of the system
interferes with structural relaxation processes in this case and
this certainly in a more pronounced manner for $\dot{\gamma}=10^{-6}$
than for $\dot{\gamma}=10^{-5}$. Therefore, the reduced stress
decays faster for the higher shear rate of $\dot{\gamma}=10^{-5}$.

The dashed lines in Fig.~\ref{fig7} are fits with compressed
exponentials. In these fits, the exponent $a_{\rm ce}$ is around
3.0 and the strain scale changes from $\delta \gamma^\star \approx
0.027$ for $N=1000$ to $\delta \gamma^\star \approx 0.012$ for
$N=10000$.  While the initial decay of $\sigma^\star$ strongly
depends on $N$, the second feature in $\sigma^\star$, the appearance
of a local maximum at $\gamma - \gamma_{\rm max} \approx 0.2$, does
not show significant finite-size effects.

\begin{figure}
\centering
\includegraphics*[width=7.5cm]{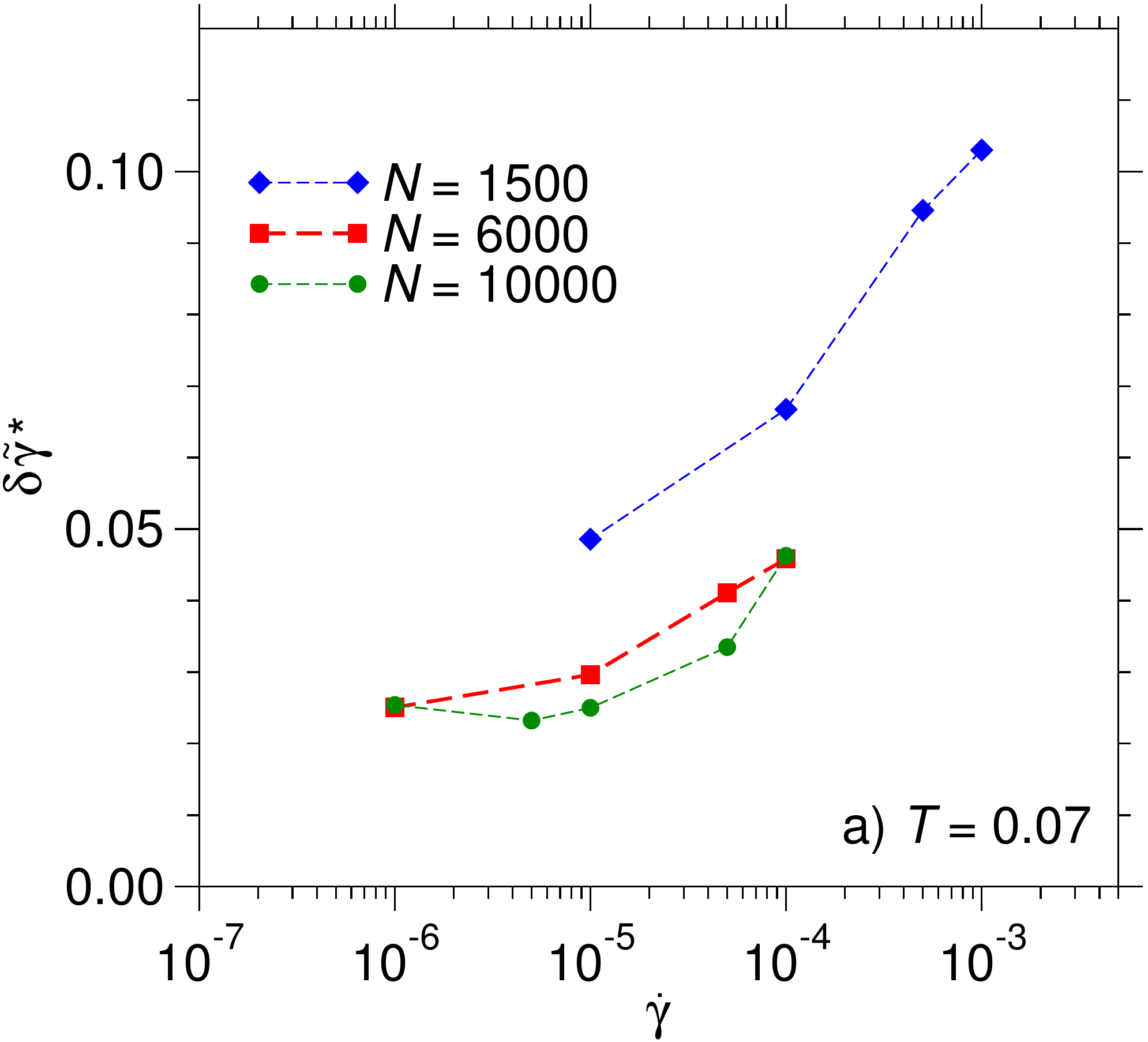}
\includegraphics*[width=7.5cm]{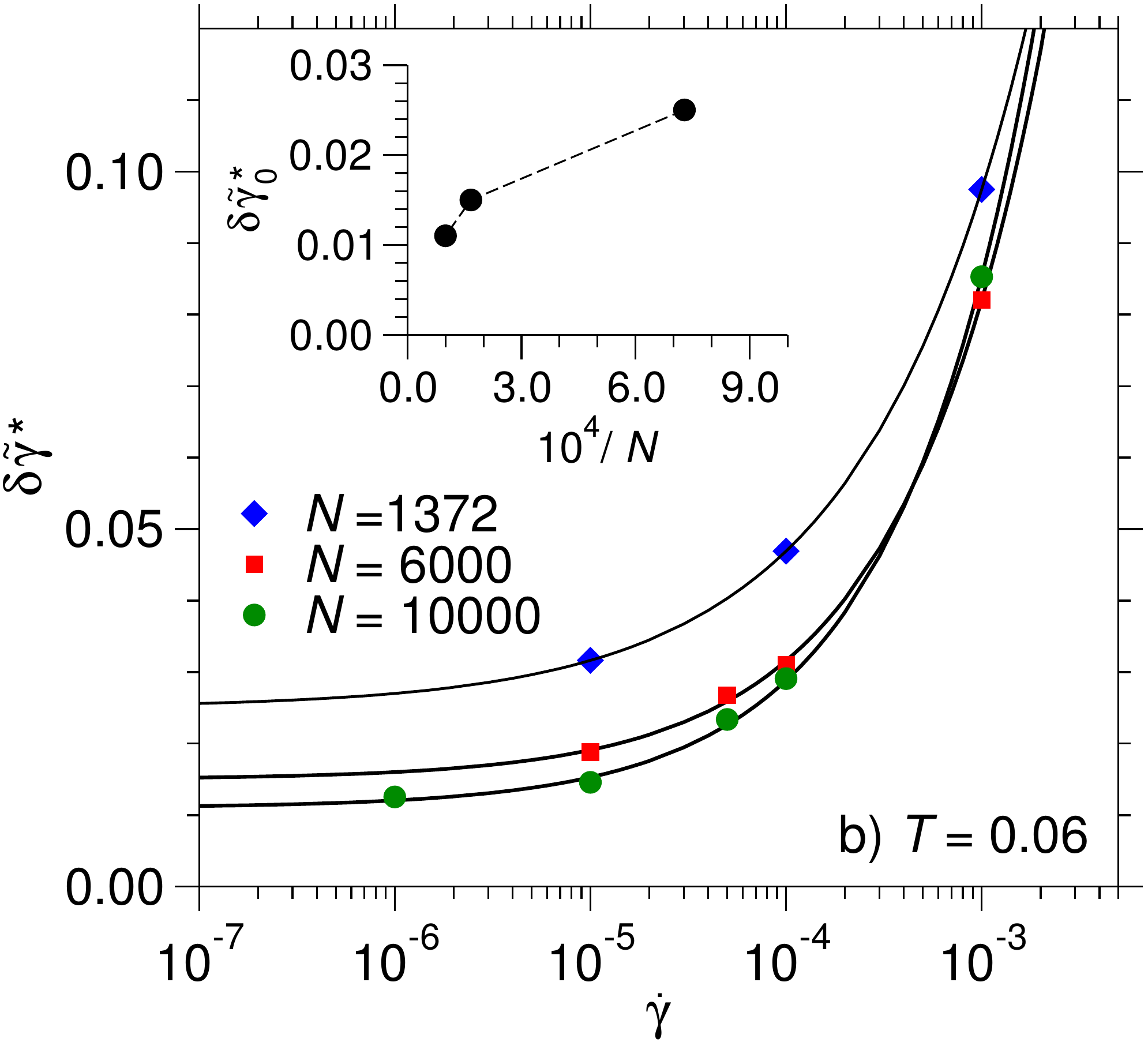}
\caption{Strain scale $\delta \tilde{\gamma}^\star$ as a function
of shear rate $\dot{\gamma}$ at the temperatures $T=0.07$ [a)] and
$T=0.06$ [b)] for different system sizes.  In b), the solid lines
are fits with the function $g(\dot{\gamma}) = \delta \tilde{\gamma}^\star_0
+ A_{\gamma} \dot{\gamma}^{c_\gamma}$ (for details see text).  The
inset shows $\delta \tilde{\gamma}^\star_0$ as a function of $N^{-1}$.
\label{fig8}}
\end{figure}
In the following, we do not use the strain scale, as directly
obtained from the fits to the compressed exponentials. In lieu
thereof, we use the value where the reduced stress, as described
by the compressed exponential, has decayed to 0.2. We denote this
quantity by $\delta \tilde{\gamma}^\star$.  Figure \ref{fig8}
displays the shear-rate dependence of $\delta \tilde{\gamma}^\star$
for $T=0.07$ in a) and $T=0.06$ in b) and different system sizes.
In the case of $T=0.07$, the transition becomes significantly sharper
for all shear rates when changing the particle number from $N=1500$
to $N=6000$. However, only small changes are observed when going
from $N=6000$ to $N=10000$ (at $\dot{\gamma}=10^{-6}$, the values
for $\delta \tilde{\gamma}^\star$ are essentially equal for the two
system sizes). This is due to the fact that the life time $\tau_{\rm
lt}$ is smaller than the time scale $\dot{\gamma}^{-1}$ for
$\dot{\gamma}\le 10^{-4}$ and thus the yielding transition interferes
with relaxation processes in the liquid.

In Fig.~\ref{fig8}b for $T=0.06$, the solid lines correspond to the
fit function $g(\dot{\gamma}) = \delta \tilde{\gamma}^\star_0 +
A_\gamma \dot{\gamma}^{c_\gamma}$, with $\delta \tilde{\gamma}^\star_0$
the estimate of $\delta \tilde{\gamma}^\star$ at zero shear rate,
$A_\gamma$ an amplitude, and $c_\gamma$ an exponent that has value
of about $0.62$ in the fits of Fig.~\ref{fig8}b. Thus, the zero
shear-rate values of the strain scale can be well estimated via a
``Herschel-Bulkley-like'' law. The inset of Fig.~\ref{fig8}b shows
$\delta \tilde{\gamma}^\star_0$ as a function of $N^{-1}$.  In the
considered range of system sizes, we observe a weak dependence of
$\delta \tilde{\gamma}^\star_0$ on system size.  The data suggests
that there might be a regime $\propto 1/N$ for large $N$, similar
to what one expects for a first-order phase transition.  However,
our data is not conclusive to support this interpretation.

But what happens at this yielding transition? And what is
the meaning of the second feature in $\sigma^\star$, i.e.~the
increase of $\sigma^\star$ up to a maximum around $\gamma-\gamma_{\rm
max} = 0.2$? Below we show that both features are connected to the
formation of shear bands. The rapid initial decay of $\sigma^\star$
is a manifestation of brittle yielding.

\begin{figure}
\centering
\includegraphics*[width=7.5cm]{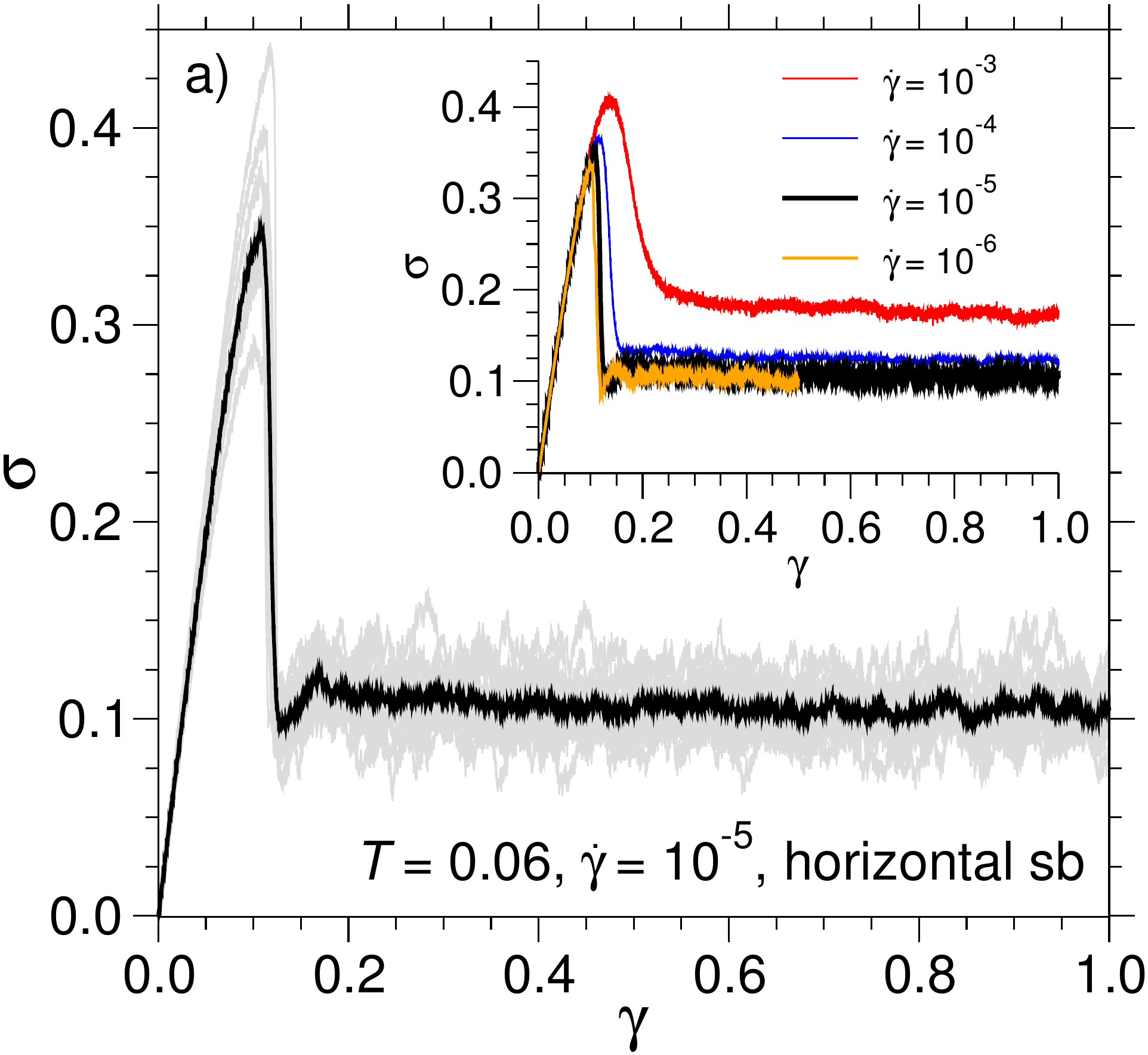}
\includegraphics*[width=7.5cm]{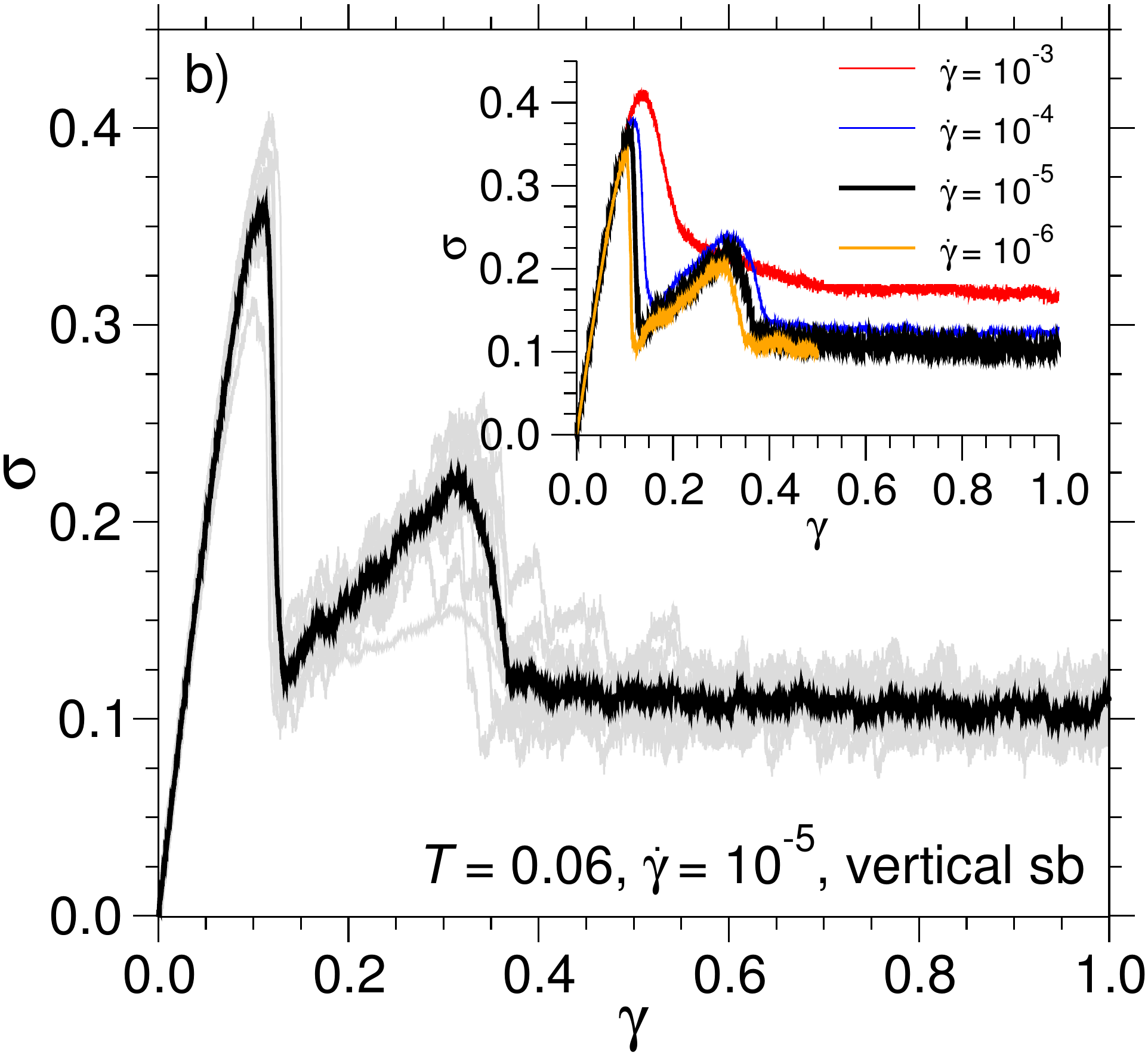}
\caption{Stress-strain relations for a) horizontal shear bands and
b) vertical shear bands at the temperature $T=0.06$ and the shear rate $\dot{\gamma}
= 10^{-5}$ in the main figures and for different shear rates in the
insets.  The grey lines in the main plots are the stress-strain relations
for the individual runs and the black lines correspond to the average
over these runs. \label{fig9}}
\end{figure}
To analyze the behavior of the system around yielding, we now
consider individual runs at the temperature $T=0.06$ and the shear
rate $\dot{\gamma} = 10^{-5}$. Among the 30 independent runs for
the systems with $N=10000$ particles, we find two types of stress-strain
relations. In both cases, we observe an initial sharp drop, indicating
brittle yielding.  However, while we see in the first type only the
initial stress drop (Fig.~\ref{fig9}a), in the second one there is
the additional increase after the first drop up to $\gamma \approx
0.32$, followed by a second drop of the stress (Fig.~\ref{fig9}b).
The first type of stress-strain relation corresponds to the formation
of a horizontal shear band, i.e.~the occurrence of a thin melted
layer with an orientation parallel to the flow direction. The second
type of stress-strain relation corresponds to the initial formation
of a vertical shear band where the melted thin layer is oriented
perpendicular to the flow direction.  Note that among the 30 runs,
we have observed horizontal and vertical shear bands in 18 and in
12 cases, respectively. In Fig.~\ref{fig9}, the grey lines
correspond to the individual runs and black ones to the average
over these runs in each case. The insets of Fig.~\ref{fig9} show
the averaged stress-strain relations for different shear rates.
For $\dot{\gamma} \le 10^{-4}$, a qualitatively similar behavior
is seen with essentially the initial stress drop getting slightly
sharper with decreasing shear rate (cf.~Fig.~\ref{fig8}).  At
$\dot{\gamma} = 10^{-3}$, however, the yielding transition is
washed out and, instead of the second maximum in the stress-strain
relation for the vertical shear bands, there is a logarithmic decay
around $\dot{\gamma} = 0.3$ (cf.~Fig.~\ref{fig6}a).

\begin{figure}
\centering
\includegraphics*[width=7.5cm]{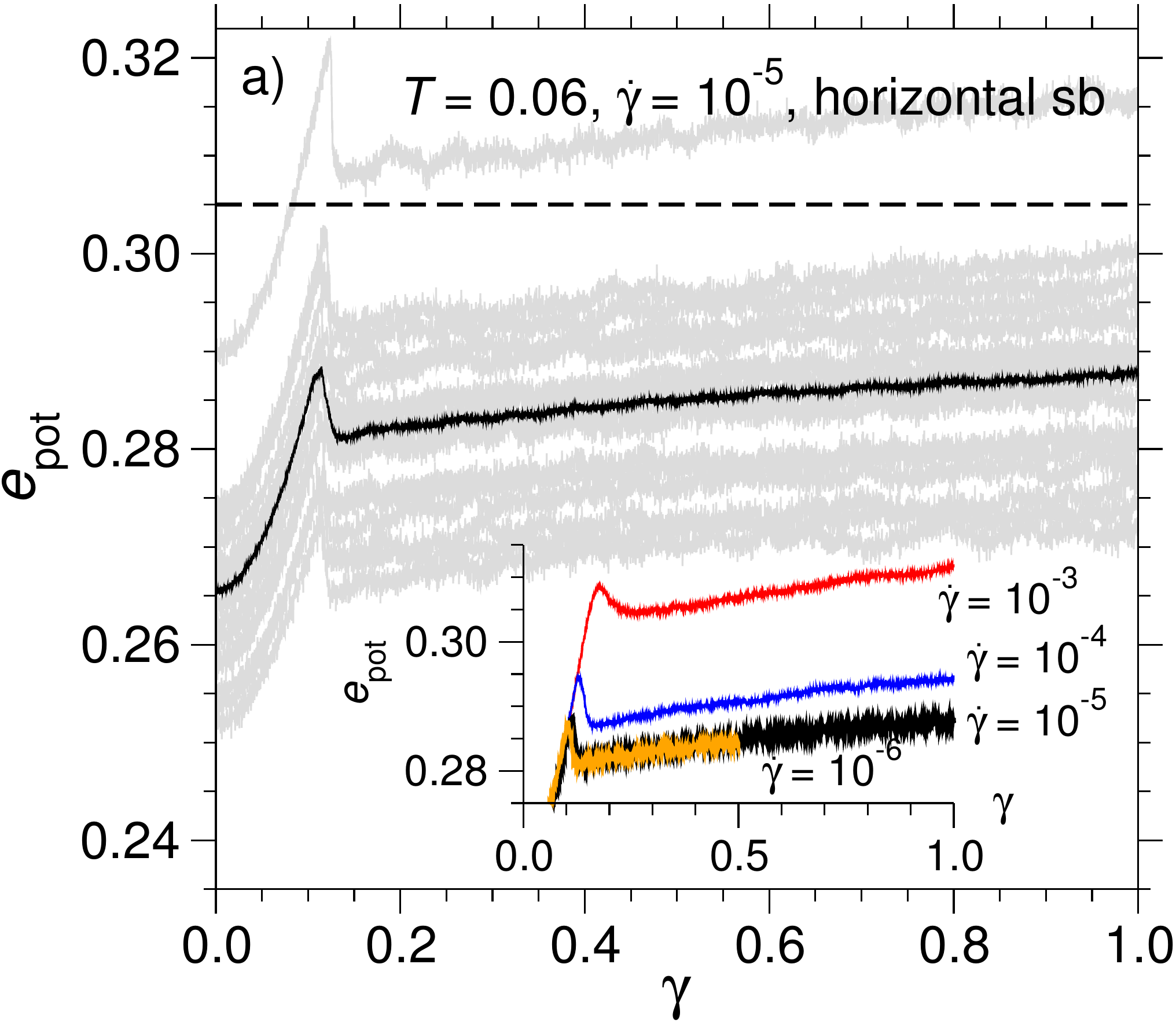}
\includegraphics*[width=7.5cm]{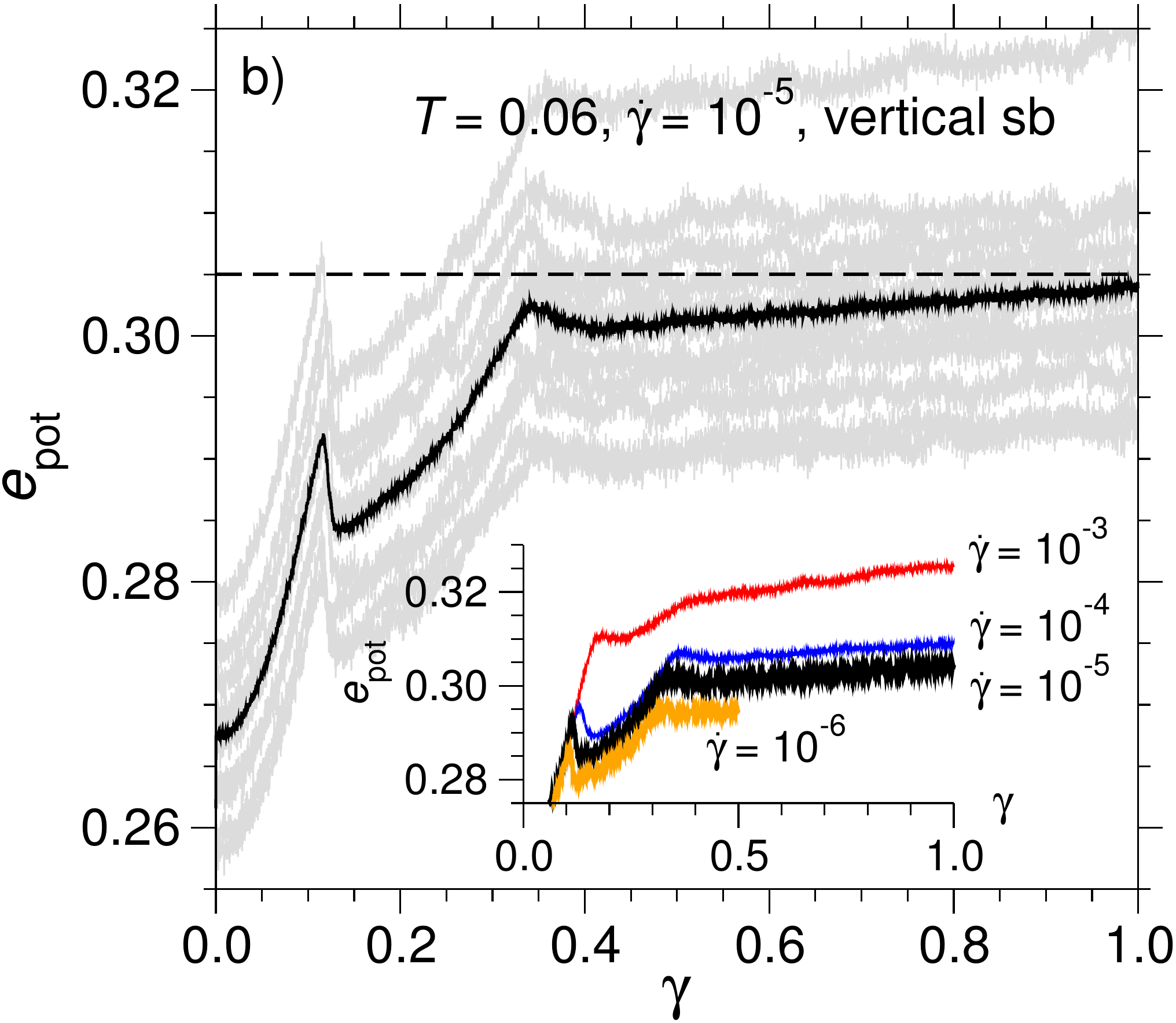}
\caption{Potential energy per particle as a function of strain for
a) horizontal shear bands and b) vertical shear bands at the
temperature $T=0.06$ and the shear rate $\dot{\gamma} = 10^{-5}$
in the main figures and for different shear rates in the insets.
The grey lines in the main plots are the stress-strain relations
for the individual runs and the black lines correspond to the average
over these runs. The horizontal dashed lines in both panels mark the average
steady-state value, $\bar{e}_{\rm pot}=0.305$. \label{fig10}}
\end{figure}
The brittle yielding is also reflected in the behavior of the
potential energy per particle, $e_{\rm pot}$, as a function of the
strain $\gamma$. In the case of the horizontal shear bands at
$T=0.06$ and $\dot{\gamma} = 10^{-5}$ (Fig.~\ref{fig10}a), there
is first a drop of $e_{\rm pot}$ at the yield point, followed by a
slow increase towards the steady-state value which is at about $\bar{e}_{\rm pot}=0.305$
(horizontal dashed line). As can be inferred from the figure, there
is a large scatter in the values of $e_{\rm pot}$ from sample to
sample.  This is due to the polydispersity of the samples. However,
the shape of the curves for the different samples is very similar
and they are essentially shifted with respect to each other. This
is also true for the behavior of $e_{\rm pot}$ vs.~$\gamma$ for the
case of the vertical shear bands (Fig.~\ref{fig10}b). Here, after
the first drop of the energy, it increases to a value which is close
to the steady-state value, then, around $\gamma \approx 0.3$
(corresponding to the local maximum in the stress-strain relation),
it slightly decreases before it increases towards the steady-state
value. For the case of the vertical shear bands, the system reaches
the steady state much faster than in the case of the horizontal
shear bands. The behavior of $e_{\rm pot}(\gamma)$ for the different
shear rates (see insets of Fig.~\ref{fig10}) is similar to that of
the corresponding stress-strain relations.

\begin{figure}[htp]
\centering
\includegraphics*[width=8.5cm]{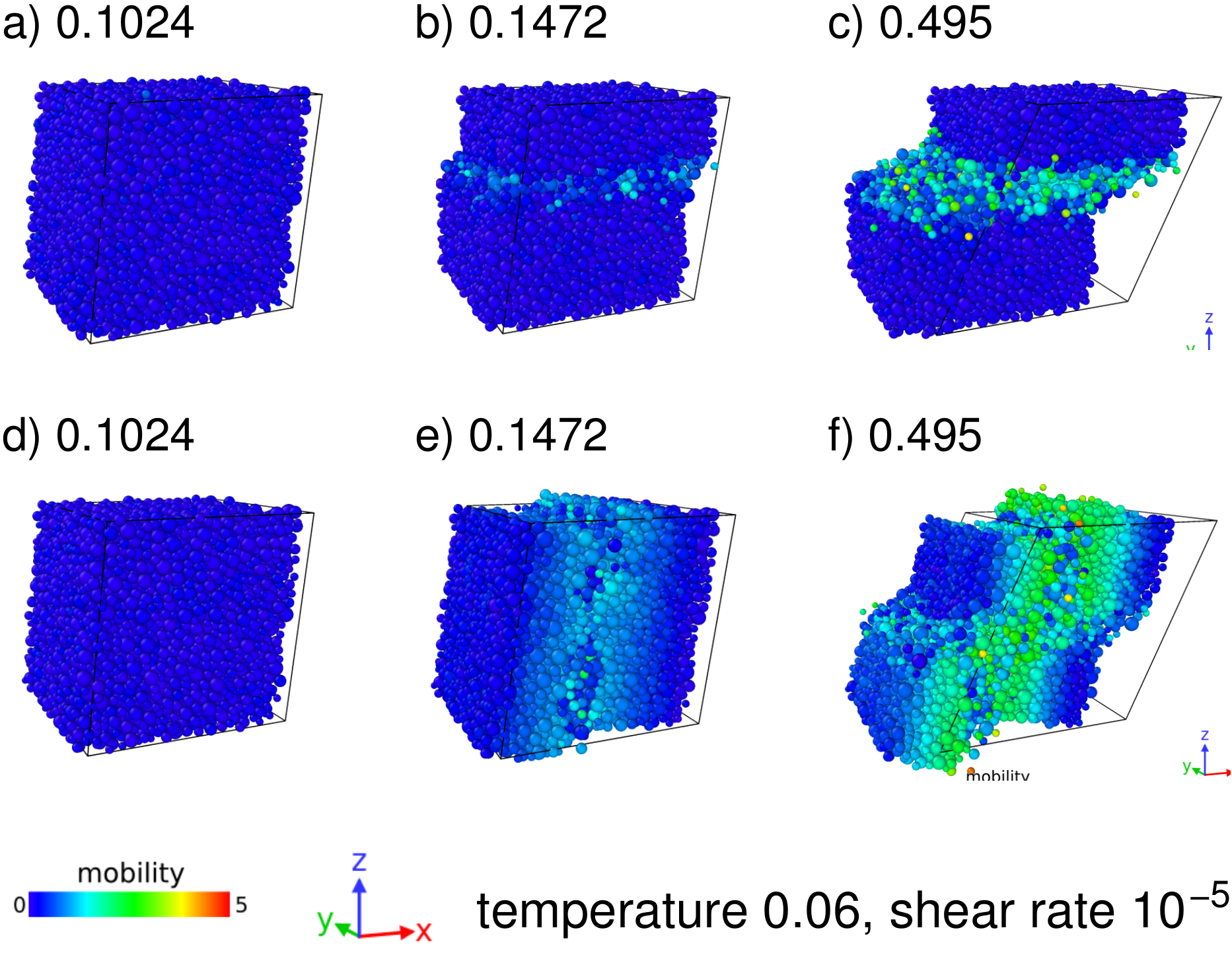}
\caption{
Mobility maps at $T=0.06$ and $\dot{\gamma}=10^{-5}$ for
a sample with the formation of a horizontal shear band
for a) $\gamma=0.1024$, b) $\gamma = 0.1472$, and c) $\gamma= 0.495$
and a sample with the formation of a vertical shear band
for the same values of $\gamma$, d) - f). 
\label{fig11}}
\end{figure}
To visualize the shear bands, mobility color maps \cite{shriva2016_1}
are computed.  To this end, we determine, for each particle $i$,
the non-averaged MSDs $\delta y_i^2(t)$ and $\delta z_i^2(t)$ in
the neutral $y$ direction and the shear-gradient $z$ direction,
respectively.  From this, we obtain the ``mobility displacement''
$\Delta_i(t) = \sqrt{\delta y_i^2(t)+\delta z_i^2}$ and we assign
a color to the magnitude of $\Delta_i$. The time origin for the
calculation of $\Delta_i$, $t=0$, corresponds to the time where the
external shear is switched on. For the snapshots in Fig.~\ref{fig11}
at different values of the strain, we have selected a sample with
a horizontal shear band [a)-c)] and one with a vertical shear band
[d)-f)], both samples are at $T=0.06$ and $\dot{\gamma} = 10^{-5}$.
At $\gamma=0.1024$, i.e.~just before the onset of plastic flow, the
system is in a homogeneously deformed state and therefore the
mobility of the particles is close to zero, as represented by the
blue color. At the strain $\gamma = 0.1472$ a horizontal shear band
has formed in the first sample (Fig.~\ref{fig11}b) and a vertical
one in the second sample (Fig.~\ref{fig11}e). In both cases, the
fluidized regions along the band are represented by particles,
colored in green. The horizontal shear band exhibits a slow growth
as a function of strain, as reflected, e.g., in a slow increase of
the potential energy of the system (cf.~Fig.~\ref{fig10}a). At
$\gamma=0.495$, the thickness of the horizontal shear band corresponds
to about 5-6\,$\sigma$, i.e.~a few liquefied layers (Fig.~\ref{fig11}c).
The behavior is different in the case of the vertical shear band.
Here, the thickness of the vertical band first increases which is
accompanied by an increase of the stress with increasing strain
(cf.~Fig.~\ref{fig9}b).  The stress drop at $\gamma \approx 0.3$
is associated with the formation of an additional horizontal shear
band which grows with increasing strain (cf.~Fig.~\ref{fig11}f).

\section{Summary and conclusions}
In summary, we have investigated the yielding behavior of a
glassforming soft-sphere model under shear.  Using molecular dynamics
(MD) simulation in combination with the swap Monte Carlo (SMC)
technique, fully equilibrated supercooled liquid samples around and
far below the critical mode coupling temperature $T_c$ were obtained.

First, these samples served as starting configurations
for simulations in the microcanonical ensemble to study how the
dynamics of the supercooled liquid changes when decreasing the temperature
from above to far below $T_c$.  In qualitative agreement with mode
coupling theory (MCT), we have seen that the reduced localization
length $\xi/\bar{d}$, as extracted from the mean squared displacement,
shows a kink at $T_c$, changing from $\xi_c/\bar{d}=0.077$ for
$T>T_c$ to a roughly linearly decreasing function for decreasing
temperature below $T_c$. Here, the critical value $\xi_c/\bar{d}$
marks the stability limit of the amorphous solid. In fact, the
decrease of $\xi$ with decreasing temperature is accompanied by
an exponential increase of a time scale $\tau_{\rm lt}$ that measures
the life time of the amorphous solid state. The Arrhenius law that we find
for the temperature dependence of $\tau_{\rm lt}$ is consistent
with the interpretation of an activated dynamics for structural
relaxation processes below $T_c$.

The gradual change of structural relaxation from a liquid-like to
a solid-like dynamics around $T_c$ is associated with a change of
the system's response to a mechanical load, in particular with
respect to the yielding of the system. In this work, we have studied
sheared supercooled liquids in a planar Couette flow geometry,
applying a constant shear rate $\dot{\gamma}$. We have shown that
the emergence of a transient amorphous solid state implies the
possibility of brittle yielding which is characterized a sharp
stress drop in the stress-strain relation. This means that around
a strain of the order of 0.1, the stress shows a sudden decrease
on a strain scale $\delta \tilde{\gamma}^\star$ much less than 0.1
(this value is found for the stress decay at yielding for temperatures
above and around $T_c$). For example, at a temperature $T=0.06$ and
a shear rate $\dot{\gamma} = 10^{-5}$, we find $\delta \tilde{\gamma}^\star =
0.014$. While at low temperatures, $T \ll T_c$, $\delta \tilde{\gamma}^\star$
significantly decreases with increasing system size, our data is
not conclusive with respect to the question whether brittle yielding
can be interpreted in terms of an underlying kinetic first-order
transition in the limit $\dot{\gamma}\to 0$.  Anyway, at finite
temperature, such an interpretation has to be taken with a grain
of salt. On the one hand, the signatures of a first-order transition
can be only seen on the time scale $\tau_{\rm lt}$ and thus for
shear rates $\dot{\gamma}$ with $\dot{\gamma}\tau_{\rm lt}\gtrsim
1$ (note that for $\dot{\gamma} \tau_{\rm lt} \gg 1$, one expects Newtonian
behavior). On the other hand, at a given temperature $T<T_c$, the
shear rate has to be small enough that the steady-state stress is
close to the apparent yield stress, as obtained from the extrapolation
to $\dot{\gamma}\to 0$ in terms of a Herschel-Bulkley law.
Thus, the time scale $\tau_{\rm lt}$ has to be very large in order
to see the signatures of a first-order transition and this is the
case for temperatures far below $T_c$.
As similar interplay of time scales has been recently found by
Shrivastav and Kahl \cite{shrivastav2021}, studying the yielding 
in a cluster crystal.

Brittle yielding is associated with horizontal or vertical shear
bands. Both types of shear bands are equally efficient to release
the stress at the yield strain. The mechanism for the formation of
such shear bands at finite temperatures and shear rates is still
not well understood, but for the transient amorphous solid states
under equilibrium conditions, as studied in this work, techniques
and theoretical frameworks can be adapted that have been previously
mainly used for athermal systems, such as an analysis of soft modes
determined from the dynamical matrix \cite{jaiswal2016, procaccia2017,
parisi2017}.  Another promising framework to investigate the yielding
transition and shear banding in amorphous solids is the analysis
in terms of non-affine displacements \cite{falk1998, lemaitre2006,
ganguly2013, ganguly2015, baggioli2021}, as recently applied to
elucidate plasticity and yielding in crystalline solids \cite{ganguly2017,
nath2018, reddy2020}. Work in this direction is in progress.

\begin{acknowledgments}
J.H. gratefully acknowledges useful discussions with Parswa Nath and
Surajit Sengupta.
\end{acknowledgments}

\end{document}